\documentclass[journal]{IEEEtran}

\usepackage[utf8]{inputenc} 
\usepackage[T1]{fontenc}    
\usepackage{hyperref}       
\usepackage{url}            
\usepackage{booktabs}       
\usepackage{amsfonts}       
\usepackage{nicefrac}       
\usepackage{microtype}      
\usepackage{xcolor}         
\usepackage{amsmath}
\usepackage{amsthm}
\usepackage{amssymb}
\usepackage{adjustbox}
\usepackage{balance}
\usepackage{enumerate}
\usepackage{enumitem}

\usepackage{algorithm}
\usepackage{algpseudocode}
\usepackage{listings}
\usepackage{cite}
\usepackage{mathtools}

\usepackage{graphicx}
\usepackage{caption}
\usepackage{subcaption}

\newtheorem{lemma}{Lemma}

\begin{document}
\title{Topological Dictionary Learning\vspace{.2cm}}

\author{Enrico Grimaldi, ~\IEEEmembership{Student Member,~IEEE,} Claudio Battiloro, ~\IEEEmembership{Member,~IEEE,}\smallskip\\ and Paolo Di~Lorenzo,~\IEEEmembership{Senior Member,~IEEE}
        \vspace{-.5cm}
\thanks{Grimaldi is with the Department of Computer, Control and Management Engineering, Sapienza University of Rome, Via Ariosto, 25, 00185 Rome, Italy. Claudio Battiloro is a Postdoctoral Fellow at the Biostatistics Department of Harvard University, 655 Huntington Ave, Boston, MA 02115, USA. 
Di Lorenzo is with the Department of Information Engineering, Electronics, and Telecommunications, Sapienza University of Rome, Via Eudossiana 18, 00184 Rome, Italy. E-mail: enrico.grimaldi@uniroma1.it, cbattiloro@hsph.harvard.edu, 
paolo.dilorenzo@uniroma1.it. This work has been supported by the SNS JU project 6G-GOALS under the EU’s Horizon program Grant Agreement No 101139232 \cite{strinati2024goal}, and by European Union under the Italian National Recovery and Resilience Plan of NextGenerationEU, partnership on Telecommunications of the Future (PE00000001- program RESTART). A preliminary version of this work was presented in \cite{battiloro2023parametric}.}}

\maketitle
\begin{abstract}
The aim of this paper is to introduce a novel dictionary learning algorithm for sparse representation of signals defined over combinatorial topological spaces, specifically, regular cell complexes. Leveraging Hodge theory, we embed topology into the dictionary structure via concatenated sub-dictionaries, each as a polynomial of Hodge Laplacians, yielding localized spectral topological filter frames. The learning problem is cast to jointly infer the underlying cell complex and optimize the dictionary coefficients and the sparse signal representation. We efficiently solve the problem via iterative alternating algorithms. Numerical results on both synthetic and real data show the effectiveness of the proposed procedure in jointly learning the sparse representations and the underlying relational structure of topological signals.
\end{abstract}

\begin{IEEEkeywords}
Topological signal processing, dictionary learning, Hodge theory, sparse coding. \vspace{-.3cm}
\end{IEEEkeywords}

\section{Introduction}
In recent years, Graph Signal Processing (GSP) has garnered significant research attention as it extends traditional signal processing to irregular domains, where signals are defined over the vertices of a graph rather than within Euclidean spaces \cite{shuman2013emerging}. This framework broadens signal processing to encompass non-metric spaces, utilizing proximity relationships to represent and analyze complex data structures. Advances in GSP have led to the development of numerous methods for effectively analyzing and processing such signals \cite{ortega2018graph}. The main feature of these processing tools is that they intrinsically depend on the graph's connectivity, which is encoded into the structure of the adopted shift operator, such as the adjacency and Laplacian matrices. From the definition of graph (shift) operator, is possible to derive graph filters and graph Fourier transforms \cite{isufi2024graph}, offering a robust framework for tasks such as spectral analysis, clustering \cite{von2007tutorial}, and sampling \cite{dilorenzo2017sampling} within this non-metric setting. However, despite their popularity and widespread use, graph-based representations are inherently limited in that they can only capture pairwise relationships \cite{agarwal2006higher, courtney2016generalized}. In complex interconnected systems, interactions often cannot be reduced to simple pairwise relationships, making graph representations incomplete and potentially inefficient. For instance, in biological networks, interactions among genes, proteins, or metabolites often involve multi-way relationships, which cannot be fully captured using graphs \cite{lambiotte2019genenetworks}. Similarly, in neuroscience, groups of neurons may activate simultaneously, reflecting higher-order relationships \cite{giusti2016two}. These limitations have catalyzed a renewed interest in extending GSP to account for multi-way relationships, giving rise to the emergent field of Topological Signal Processing (TSP) \cite{barbarossa2020topological,schaub2021signal}. 

TSP extends signal processing to domains defined over combinatorial topological spaces, such as simplicial complexes and cell complexes. Both these domains are combinatorial topological spaces able to encode multi-way and hierarchical relationships, enabling sophisticated adjacency schemes among \textit{cells} (nodes or groups of nodes) beyond standard graph adjacency. Informally, a cell complex can be seen as an augmented graph in which nodes, edges, and induced cycles are cells. While nodes are adjacent if connected by an edge, edges can be \textit{lower neighbors} if sharing a node or \textit{upper neighbors} if part of the same induced cycle.
Cell complexes also come with a rich algebraic description, but there is large space for exploration to different combinatorial topological spaces, based on the specific computational and modeling needs \cite{robinson2014topological}. This transition to topological domains, building upon the advancements in GSP, has already made substantial contributions \cite{isufi2024topological}, spanning from the introduction of FIR filters for simplicial and cell complex signals \cite{SCF_9893391, CellModels}, to the definition of a generalized Laplacian for embedding simplicial complexes into traditional graphs \cite{ji2022signal}, as well as the introduction of models like the self-driven graph Volterra model \cite{coutino2020self}, capable of capturing higher-order interactions in network data. Moreover, TSP has also influenced the world of machine learning and data science: many researchers introduced Topological Data Analysis techniques for extracting features to be fed to classical Machine Learning \cite{TDA_ML1} or Deep Learning models \cite{Persformer}. The rising field of Topological Deep Learning couples TSP techniques with neural models, through the extension of the message-passing framework of Graph Neural Networks \cite{GNN2} to higher-order topological domains \cite{papillon2023architectures}, or through latent topology inference \cite{battiloro2023latent}.\\ 
\noindent \textbf{Motivation and Related Works.}
Many real-world signals are compressible, meaning that they admit some sparse representation over a certain basis. Consequently, a fundamental problem in signal processing is \textit{sparse representation} \cite{rioul1991wavelets}, whose aim is designing overcomplete dictionaries of atoms (i.e. frames) that can represent signals as linear combinations of only a few atoms in the dictionary \cite{foucartCS}. This approach often implies the definition of \textit{analytical dictionaries}, i.e., structured atoms based on mathematical modeling that can be designed starting from the given domain and assuming a certain signal class. For instance, in the Euclidean domain, several techniques such as the JPEG for image signals \cite{hudson2018jpeg}, and the MPEG and MP3 for video/audio signals \cite{priji2overview}, rely on dictionaries based on Fourier transform, wavelets, or curvelets for compression. Another possibility is to infer the dictionary structure from available training signals, a procedure known as \textit{dictionary learning} (DL) \cite{tovsic2011dictionary}, \cite{forero2014prediction}. Renowned DL approaches generally fall into three main categories \cite{tovsic2011dictionary}: i) probabilistic estimation methods \cite{engan1999method}, ii) clustering or vector quantization \cite{schmid2004dictionary,aharon2006k}, iii) domain-informed parametric dictionaries \cite{mailhe2008shift,sallee2002learning}. Generally, analytic dictionaries are faster to implement but vulnerable to model mismatching; whereas, learnable dictionaries are often robust to different signal classes, but have larger complexity due to the required training phase. 


Localized dictionaries for sparsely representing graph-based data have been extensively studied, see, e.g., \cite{hammond2011wavelets,tsitsvero2016signals,behjat2016signal,shuman2020localized}. The seminal work in \cite{hammond2011wavelets} introduced a computationally efficient method for constructing invertible spectral graph wavelet transforms using the graph Laplacian, enabling localized analysis across diverse problem domains. In \cite{tsitsvero2016signals}, the authors proposed graph Slepians—a class of graph signals maximally concentrated on a specific vertex set while perfectly localized within a spectral band. The study in \cite{behjat2016signal} developed a tight frame design for graph signals, adapting Meyer-type kernels to align with the energy distribution of specific signal classes, seamlessly integrating graph topology with signal features. Lastly, \cite{shuman2020localized} provides an excellent overview of localized spectral graph filter frames, constructed by localizing patterns (i.e., spectral filters) to various regions of the graph.  In this context, several approaches to graph-based dictionary learning have been proposed as well \cite{thanou2014learning,yankelevsky2016dual,vincent2021online,boukrab2024online}. In \cite{thanou2014learning}, a parametric dictionary learning algorithm was introduced for sparse graph signal representation, integrating data adaptation, graph localization, and computational efficiency. The work in \cite{yankelevsky2016dual} proposed a dictionary learning method that jointly exploits graph topology and data manifold structure, ensuring smoothness in dictionary atoms and sparse representations while learning an adaptive graph Laplacian. In \cite{vincent2021online}, an online Graph Dictionary Learning approach was presented, utilizing Gromov-Wasserstein divergence to model graphs as convex combinations of dictionary atoms, facilitating efficient representation, embedding, and subspace tracking. Lastly, \cite{boukrab2024online} introduced a Laplacian-Regularized Dictionary Learning framework, combining dictionary and graph structure learning through Proximal Alternating Linearized Minimization, offering both batch and online optimization variants.

Considering more general topological domains going beyond graphs, the literature related to sparse signal representation is much more scarce. Some works have proposed analytical dictionaries for topological signals \cite{barbarossa2020topological,roddenberry2022hodgelets,battiloro2023topological}. In fact, generalizing what has been done for graph signals, a natural basis for signal representation is given by the topological Fourier modes, i.e., the eigenvectors of Hodge Laplacians \cite{schaub2020random,barbarossa2020topological}. However, since Fourier modes are generally non-sparse and thus inefficient for representing localized signals, the work in \cite{roddenberry2022hodgelets} proposed a family of wavelets for simplicial signals, respecting the Hodge decomposition. Also, the work in \cite{battiloro2023topological} introduced topological Slepians, i.e., a class of signals that are maximally concentrated on the topological domain and perfectly localized on the spectral domain. However, to the best of our knowledge, strategies to learn dictionaries for sparse representation of signals defined over more general topological domains are still missing in the current literature, and represent the main objective of investigation in this work.\\
\noindent \textbf{Contributions.} In this paper, leveraging formal principles from algebraic topology \cite{grady2010discrete}, we propose a novel class of learnable dictionaries for topological signals that are computationally efficient and explicitly incorporate topological structures. Building upon our preliminary study \cite{battiloro2023parametric} and extending the graph-based approach in \cite{thanou2014learning} to regular cell complexes, we employ Hodge theory to embed the underlying geometric structure into the dictionaries. Specifically, the dictionaries are designed as concatenations of sub-dictionaries, parameterized as polynomials of the Hodge Laplacians that leads to localized spectral topological filter frames. This design simultaneously exploits information redundancy through overparameterization, leverages the symmetry properties inherent in cell complex filtering, and offers provable bounds of the resulting frame operators. We consider different parametrizations of Hodge-aware topological dictionaries, leading to a trade-off between performance and complexity. The topological domain itself is only partially known: We assume perfect knowledge of the lower connectivity within the cell complex representing the data, while the upper Laplacian structure is inferred through a task-driven learning process designed to optimize the underlying domain for enhanced performance in sparse representation. The problem is mathematically cast as the joint optimization of (upper) topological structure, dictionary parameters, and sparse representation to minimize a data fitting term under topological constraints on the dictionary structure. To cope with the complexity of the resulting non-convex problem, we introduce an efficient alternating optimization procedure that proceeds iteratively by decomposing the formulation into sub-problems that are easier to solve and control. Finally, we show the performance of the proposed methodology over both synthetic data and real traffic flows, illustrating the compression capabilities, the topology inference performance, and the comparison with other available approaches. To summarize, the main contributions of this paper are the following:\vspace{-.05cm}
\begin{itemize}[leftmargin=*]
\item We introduce a novel class of localized spectral topological filter frames, designed to tailor spectral filter patterns to specific regions of the cell complex. Furthermore, we derive theoretical frame bounds for the proposed topological dictionaries, guaranteeing their robustness and applicability.
\item We propose the first method for topological dictionary learning, which jointly learns the upper Laplacian structure of the cell complex domain, the parameters of the topological filters composing the dictionary, and the sparse representation from available training signals.
\item We propose two iterative optimization procedures to address the non-convex dictionary learning problem by alternately solving simpler sub-problems. The first procedure employs orthogonal matching pursuit (OMP) for sparse coding, quadratic programming (QP) for optimizing the dictionary coefficients, and a greedy search algorithm to identify the cells present in the complex. The second procedure can be regarded as a relaxed variant of the first, in which the greedy search step is replaced by a proximal gradient descent algorithm for topology inference.
\item We demonstrate the advantages of the proposed methodologies on both synthetic datasets and real-world traffic flows. Our approach outperforms existing methods that rely solely on graph-based information or analytical topological dictionaries, showcasing its superior effectiveness and versatility.
\end{itemize}

\textbf{Outline.} This paper is structured as follows: Section II reviews essential mathematical tools from topological signal processing, establishing the necessary foundation. Section III introduces localized spectral topological filter frames and their mathematical properties. Section IV formulates the proposed topological dictionary learning problem, while Section V presents its algorithmic solutions. Section VI provides numerical results demonstrating the effectiveness of the proposed methods. Finally, Section VII concludes with a summary of findings and insights. 

\noindent \textbf{Notation.} Scalar, column vector, and matrix variables are indicated by plain letters $a$, bold lowercase letters $\mathbf{a}$, and bold uppercase letters $\mathbf{A}$, respectively. $a(i)$ is the $i$-th element of vector $\mathbf{a}$, $[\mathbf{A}]_{i,j}$ is the $(i,j)$-th element of $\mathbf{A}$, $[\mathbf{A}]_i$ is the $i$-th row of $\mathbf{A}$, $[\mathbf{A}]^i$ is the $i$-th column of $\mathbf{A}$, and $\lambda_{MAX}(\mathbf{A})$ denotes the largest eigenvalue of $\mathbf{A}$.  $\mathbf{I}$ is the identity matrix, $\mathbf{1}$ is the vector of all ones, and $\mathbf{1}_{\mathcal{S}}$ is a vector with ones at the indices in $\mathcal{S}$ and zeros elsewhere. $\textrm{im}(\cdot)$, $\textrm{ker}(\cdot)$, and $\text{supp}(\cdot)$ denote the image, the kernel, and the support of a matrix, respectively; $\oplus$ is the direct sum of vector spaces, and $\otimes$ denotes the Kronecker product. $\{\mathbf{a}_k\}_{k=1}^K$ and $\{\mathbf{A}_k\}_{k=1}^K$ are the collection of $K$ vectors and matrices, respectively. Other specific notation is defined along the paper, if needed.

\section{Background on Topological Signal Processing}

In this section, we review the necessary topological signal processing tools. Our focus is on regular cell complexes, fairly general combinatorial topological spaces able to model a wide class of higher-order hierarchical interactions \cite{grady2010discrete}. 

\noindent\textit{\textbf{Definition 1 (Regular Cell Complex).}} A {\it regular cell complex}  is a topological space $\mathcal{X}$ together with a partition $\{\mathcal{X}_{\sigma_i}\}_{\sigma_i \in \mathcal{P}_{\mathcal{X}}}$ of subspaces $\mathcal{X}_{\sigma_i}$ of $\mathcal{X}$ called \textit{cells}, where $\mathcal{P}_{\mathcal{X}}$ is the indexing set of $\mathcal{X}$, such that \cite{hansen2019toward}:

\begin{enumerate}
    \item For each $c$ $\in$  $\mathcal{X}$, every sufficient small neighborhood of $c$ intersects finitely many $\mathcal{X}_{\sigma_i}$;  
    \item For all $\sigma_i$,$\sigma_j$ we have that $\mathcal{X}_{\sigma_i}$ $\cap$ $\overline{\mathcal{X}}_{\sigma_j}$ $\neq$ $\varnothing$ iff $\mathcal{X}_{\sigma_i}$ $\subseteq$ $\overline{\mathcal{X}}_{\sigma_j}$, where $\overline{\mathcal{X}}_{\sigma_j}$ is the closure of the cell $\sigma_j$;
    \item Every $\mathcal{X}_{\sigma_i}$ is homeomorphic to $\mathbb{R}^{k}$ for some $k$;
    \item For every $\sigma_i$ $\in$ $\mathcal{P}_{\mathcal{X}}$ there is a homeomorphism $\phi$ of a closed ball in $\mathbb{R}^{k}$ to $\overline{\mathcal{X}}_{\sigma_i}$ such that the restriction of $\phi$ to the interior of the ball is a homeomorphism onto $\mathcal{X}_{\sigma_i}$.
\end{enumerate}

Condition 2 implies that the indexing set $\mathcal{P}_{\mathcal{X}}$ has a poset structure, given by $\sigma_i$ $\leq$ $\sigma_j$ iff $\mathcal{X}_{\sigma_i}$ $\subseteq$ $\overline{\mathcal{X}}_{\sigma_j}$, and we say that $\sigma_i$ \textit{bounds} or \textit{is incident to} $\sigma_j$. This is known as the \textit{face poset} of $\mathcal{X}$. The regularity condition 4 implies that all of the topological information about $\mathcal{X}$ is encoded in the poset structure of $\mathcal{P}_{\mathcal{X}}$. Then, a regular cell complex can be identified with its face poset. For this reason, from now on we will indicate the cell $\mathcal{X}_{\sigma_i}$ with its corresponding face poset element $\sigma_i$. The dimension $\textrm{dim}(\sigma_i)$ of a cell $\sigma_i$ is $k$, we call it a $k-$cell and denote it with $\sigma_i^k$. The order of a cell complex is the largest dimension of any of its cells and we denote an order $K$ regular cell complex with $\mathcal{X}_K$. Regular cell complexes can be described via a boundary relation consistent with the partial order in Definition 1.

\noindent\textit{\textbf{Definition 2 (Boundary Relation).}} We write  $\sigma_i$ $\prec$ $\sigma_j$ iff $\dim({\sigma_i})$ $\leq$ $\dim({\sigma_j})$ and there is no $\sigma_p$ s.t. $\sigma_i$ $\leq$ $\sigma_p$ $\leq$ $\sigma_j$.

 In other words, Definition 2 states that the boundary of a cell $\sigma_j^k$ of dimension $k$ is the set of all cells of dimension $k-1$ bounding $\sigma_j^k$.  
 
 \noindent\textit{\textbf{Definition 3 ($k$-skeleton).}}The {\it k-skeleton} of a regular cell complex $\mathcal{X}$ is the subcomplex of $\mathcal{X}$ consisting of cells of dimension at most $k$.
\begin{figure}
    \centering
    \includegraphics[width=1.05\linewidth]{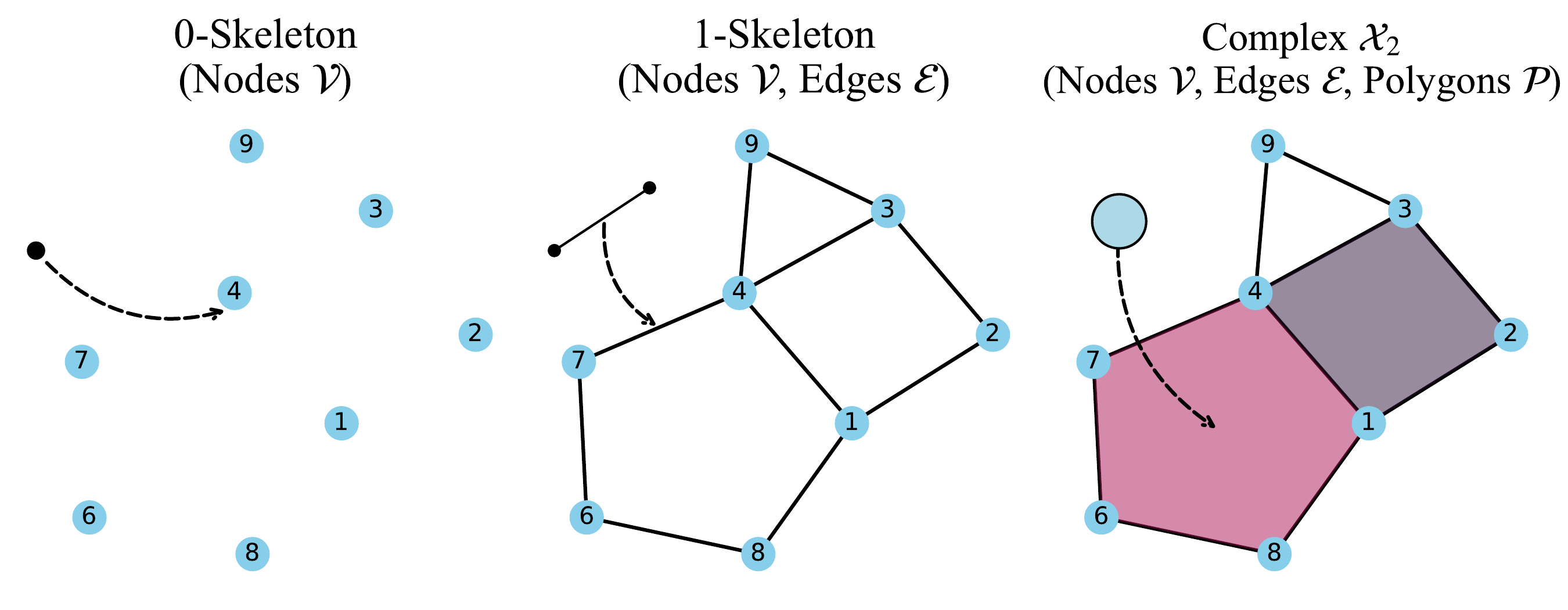}
    \caption{Hierarchical gluing, starting from a set of nodes and deriving edges (center) and polygons (right) as $1$- and $2$-cells. 
    }
    \label{fig:Gluing}
\end{figure}

 \noindent\textbf{Cell Complexes via Hierarchical Gluing.} Very often, regular cell complexes are constructed through a \textit{hierarchical
gluing procedure} starting from a finite set of \textit{vertices} $\mathcal{V}$ \cite{hansen2019toward}. In this case, a cell complex of order one $\mathcal{X}_1$ is equivalent to a graph, as it is made by the set $\mathcal{V}$ of 0-cells --the vertices-- and a  set $\mathcal{E}$ of 1-cells --pairs of vertices, i.e., the \textit{edges}--. There are several possible ways to define 2-cells --that we refer to as \textit{polygons} and we collect in a set $\mathcal{P}$-- to obtain a cell complex of order two $\mathcal{X}_2$ \cite{bodnar2021weisfeiler}. For example, polygons can be attached to the/a set/subset of simple/induced cycles of the underlying graphs, i.e. the 1-skeleton of $\mathcal{X}_2$, as shown in Fig. \ref{fig:Gluing}. Cells of order greater than 2 can be defined using the same gluing procedure \cite{sardellitti2022cell,grady2010discrete}, but there is usually little interest in them. 

The  set of $k$-cells in $\mathcal{X}_{K}$ is denoted by ${\cal D}_{k} := \{\sigma_{i}^k: \sigma_{i}^k \in \mathcal{X}_{K}\} $, with $|{\cal D}_{k}| = N_k$. In the case of a cell complex  of order two $\mathcal{X}_2$, we then have  ${\cal D}_{0}={\cal V}$, ${\cal D}_{1}={\cal E}$, and ${\cal D}_2={\cal P}$.

\noindent\textit{\textbf{Definition 4 (Topological Signals).}} 
A $k$-topological signal $\mathbf{y}_k$ over a $K$-th order regular cell complex $\mathcal{X}_K$ is defined as a collection of mappings from the set of all $k$-cells contained in the complex to real numbers:
\begin{equation}
\mathbf{y}_k = [y_k(\sigma^k_1), \ldots, y_k(\sigma^k_i), \ldots, y_k(\sigma^k_{N_k})] \in \mathbb{R}^{N_k},
\end{equation}
where $\mathbf{y}_k : \mathcal{D}_k \rightarrow \mathbb{R}$. In the case of a cell complex of order two $\mathcal{X}_2$, we then have node, edge, and polygon signals $\mathbf{y}_0$, $\mathbf{y}_1$, $\mathbf{y}_2$, respectively, determined by the corresponding mappings:
\begin{equation}
    y_{0}: {\cal V} \rightarrow \mathbb{R} , \qquad y_{1}: {\cal E} \rightarrow \mathbb{R} , \qquad y_{2}: {\cal P} \rightarrow \mathbb{R}.
\end{equation}
\noindent\textbf{Algebraic Representation.} The structure of a cell complex ${\cal X}_{K}$  is fully described by the set of its incidence matrices $\mathbf{B}_{k}$, $k=1, \ldots, K$, given a reference orientation \cite{goldberg2002combinatorial}. Although orienting cells is not mathematically trivial, it is only a "bookkeeping matter" \cite{roddenberry2022cellsp}. For this reason, here we assume that a reference orientation of the complex is given, and detailed explanations can be found in \cite{sardellitti2022cell,roddenberry2022cellsp}. The incidence matrices reflect the boundary relation from Definition 2, i.e., the non-zero entries of $\mathbf{B}_{k}$ establish which $k$-cells are incident to which $(k-1)$-cells. We use the notation $\sigma^{k-1}_i \sim \sigma^k_j$ to indicate two cells with the same orientation,  and  $\sigma^{k-1}_i \sim \sigma^k_j$ to indicate that they have opposite orientation. Mathematically,  the entries of $\mathbf{B}_{k}$ are defined as follows:
  \begin{equation} \label{inc_coeff}
  \big[\mathbf{B}_{k} \big]_{i,j}=\left\{\begin{array}{rll}
  0, & \text{if} \; \sigma^{k-1}_i \not\prec \sigma^k_j \\
  1,& \text{if} \; \sigma^{k-1}_i \prec \sigma^k_j \;  \text{and} \; \sigma^{k-1}_i \sim \sigma^k_j\\
  -1,& \text{if} \; \sigma^{k-1}_i \prec \sigma^k_j \;  \text{and} \; \sigma^{k-1}_i \not\sim \sigma^k_j\\
  \end{array}\right. .
  \end{equation}
In the case of a cell complex of order two $\mathcal{X}_{2}$, we have two incidence matrices $\mathbf{B}_{1} \in \mathbb{R}^{N_0 \times N_1}$ and  $\mathbf{B}_{2} \in \mathbb{R}^{N_1 \times N_2}$. From the incidence information, we can build the Hodge Laplacian matrices \cite{goldberg2002combinatorial}, of order $k=0, \ldots, K$, as follows:
\begin{align}
&\mathbf{L}_{0}=\mathbf{B}_{1}\mathbf{B}_{1}^T,\label{Laplacian0}\\
&\mathbf{L}_{k}=\underbrace{\mathbf{B}_k^{T}\mathbf{B}_{k}}_{\mathbf{L}_k^{(d)}}+\underbrace{\mathbf{B}_{k+1}\mathbf{B}_{k+1}^T}_{\mathbf{L}_k^{(u)}}, \; k=1, \ldots, K-1, \label{Laplaciank}\\
&\mathbf{L}_{K}=\mathbf{B}_{K}^T\mathbf{B}_{K}.\label{LaplacianK}
\end{align}
\begin{figure}
    \centering
    \includegraphics[width=1.05\linewidth]{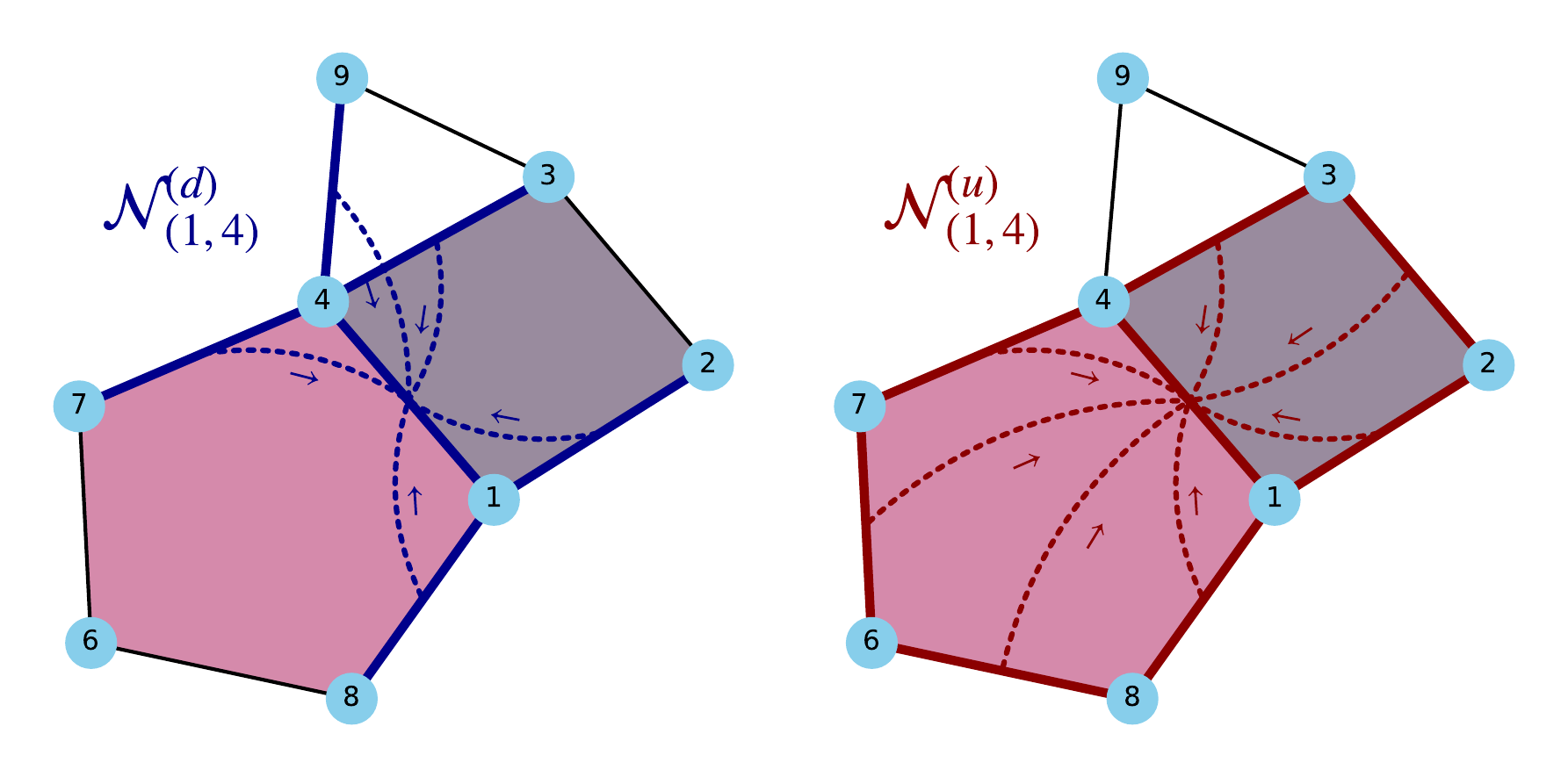}
    \caption{In the left panel, the blue edges are lower-adjacent to the edge \( (1,4) \), while the right panel displays the red edges, which are upper-adjacent to the same edge. We denote the lower and upper neighborhoods of the edge \( (1,4) \) by \(\mathcal{N}^{(d)}_{(1,4)}\) and \(\mathcal{N}^{(u)}_{(1,4)}\), respectively.}
    \label{fig:adjacency}
\end{figure}
\noindent\textbf{Cell Complex Adjacencies.} All Laplacian matrices in \eqref{Laplaciank}
of intermediate order, i.e. $k=1, \ldots, K-1$, contain two terms: the first term $\mathbf{L}^{(d)}_k$, also known as \textit{lower Laplacian}, encodes \textit{the lower adjacency} of $k$-cells; the second term $\mathbf{L}_k^{(u)}$, also known as \textit{upper Laplacian}, encodes \textit{the upper adjacency} of $k$-cells. In particular, we say that two $k$-cells are upper adjacent if they are both incident to the same $k+1$-cell, while they are lower adjacent if they are both bounded by the same $k-1$-cell. Thus, for example, two edges are lower adjacent if they share a common vertex, whereas they are upper adjacent if they are sides of a common polygon, as we show in Fig. \ref{fig:adjacency}. Note that the vertices of a graph can only be upper adjacent if they are incident to the same edge. This is why the Laplacian $\mathbf{L}_0$ contains only one term, and it is the usual graph Laplacian.

\noindent\textbf{Hodge Decomposition} Hodge Laplacians admit a Hodge decomposition, stating that the signal space associated with each cell of order $k$ can be decomposed as the direct sum of the following three orthogonal subspaces \cite{lim2020hodge}:
\begin{equation} \label{hodge_spaces}
\mathbb{R}^{N_{k}} = \text{im}(\mathbf{B}_{k}^T\big) \oplus \text{im}\big(\mathbf{B}_{k+1}\big) \oplus \text{ker}\big(\mathbf{L}_{k}\big).
\end{equation}
The dimensions of $\text{ker}(\mathbf{L}_k)$ are also known as {\it Betti numbers} of order $k$:  $\text{ker}(\mathbf{L}_0)$ is the number of connected components of the underlying graph, $\text{ker}(\mathbf{L}_1)$ is the number of holes, $\text{ker}(\mathbf{L}_2)$ is the number of cavities, and so on \cite{eckmann1944harmonische}. As a consequence of \eqref{hodge_spaces}, every signal $\mathbf{y}_{k}$ of order $k$ can be decomposed as:
\begin{equation}
\label{prop:hodge_dec}
\mathbf{y}_{k}=\mathbf{B}_{k}^T\, \mathbf{y}_{k-1} +\mathbf{B}_{k+1}\, \mathbf{y}_{k+1}+\widetilde{\mathbf{y}}_{k}.
\end{equation}
In the case $k=1$, we refer to $\mathbf{B}_{k}^T\, \mathbf{y}_{k-1}$, $\mathbf{B}_{k+1}\, \mathbf{y}_{k+1}$ and  $\widetilde{\mathbf{y}}_{k}$ as the irrotational, solenoidal, and harmonic component of $\mathbf{y}_{k}$, respectively. The Hodge decomposition reserves an intuitive interpretation for edge signals $\mathbf{y}_1$, as described in \cite{barbarossa2020topological, sardellitti2022cell}. 

\noindent\textbf{Spectral Domain and Filters.} When considering graphs, the \textit{graph Fourier transform} \cite{sardellitti2017graph, sandryhaila2013discrete} is defined as the projection of node signals onto the basis of the eigenvectors of the graph Laplacian in \eqref{Laplacian0}. It is possible to define a cell complex Fourier transform following similar arguments \cite{sardellitti2022cell}.

\noindent\textit{\textbf{Definition 5 (Topological Fourier Transform).}}
Denoting the eigendecomposition of the $k$-th order Hodge Laplacian $\mathbf{L}_k = \mathbf{U}_k \mathbf{\Lambda}_k \mathbf{U}_k^T$, the Topological Fourier Transform (TFT) of a $k$-topological signal $\mathbf{y}_k$ is the projection of the signal onto the basis of the eigenvectors of $\mathbf{L}_k$ \cite{sardellitti2022cell}, i.e.,
\begin{equation}\label{eq:tft}
\hat{\mathbf{y}}_k = \mathbf{U}^T_k \mathbf{y}_k.
\end{equation}
We refer to the eigenvalue domain of the TFT as the \textit{frequency domain} (or spectrum). A consequence of the Hodge decomposition in \eqref{prop:hodge_dec} is that the eigenvectors belonging to $\mathbf{im}(\mathbf{L}_k^{(d)})$ are orthogonal to those belonging to $\mathbf{im}(\mathbf{L}_k^{(u)})$ for all $k=1,\cdots,K-1$. The existence of a well-defined frequency domain enables the possibility of filtering topological signals.

\noindent\textit{\textbf{Definition 6 (Cell Complex FIR Filters).}}
A Cell Complex FIR Filter $\mathbf{H} \in \mathbb{R}^{N \times N}$ for a $k$-th order topological signal $\mathbf{y}_k$ is a polynomial of the $k$-th Hodge Laplacian $\mathbf{L}_k$ \cite{roddenberry2022cellsp}, i.e.:
\begin{equation}
\label{eq:joint_filter}
    \mathbf{H}\mathbf{y}_k = \sum_{j=0}^J h_j (\mathbf{L}_k)^j \mathbf{y}_k,
\end{equation}
where $J$ is a positive integer and $\mathbf{h}=\{h_j\}_{j=0}^{J} \in \mathbb{R}^{(J+1)}$ is the vector containing the filter coefficients.  We can explicitly leverage the Hodge decomposition of the $k$-th order Laplacian into its upper and lower components to define the more flexible version of Cell Complex FIR Filter introduced in \cite{sardellitti2022cell}:
\begin{equation}
\label{eq:sep_filter}
    \mathbf{H}\mathbf{y}_k = \left( \sum_{j=1}^J \Big( h^{(d)}_{j} (\mathbf{L}^{(d)}_k)^j + h^{(u)}_{j} (\mathbf{L}^{(u)}_k)^j \Big)+h^{(id)}\mathbf{I}\right)\mathbf{y}_k,
\end{equation}
where three different sets of filter coefficients $\mathbf{h}^{(d)}=[h_1^{(d)},\cdots,h_J^{(d)}] \in \mathbb{R}^{J}$, $\mathbf{h}^{(u)}=[h_1^{(u)},\cdots,h_J^{(u)}] \in \mathbb{R}^{J}$, and $h^{(id)}$ are used to filter the signal based on its rotational, solenoidal, and harmonic components, respectively. 

 Please notice that the Cell Complex FIR filters in \eqref{eq:sep_filter} are \textit{localized} shift operations that replace a signal value at each $k$-cell with a weighted sum (by the filter coefficients) of the linear combinations of the signal values over the lower and upper $j$-hop adjacencies of the $k$-cell, encoded in the $j$-th powers $(\mathbf{L}^{(d)}_k)^j$ and $(\mathbf{L}^{(u)}_k)^j$, respectively, up to order $J$. In the next section, we will leverage this feature to define localized topological spectral frames, being key objects for our dictionary learning framework that we will present in Section \ref{sec:param_dic_learn}.
 
\section{Localized Topological Spectral Frames}\label{sec:loc_spec_frames}
In this section, we leverage the Topological Fourier Transform (TFT) from \eqref{eq:tft} to build localized, overcomplete, and parametric dictionaries guaranteed to be frames for the corresponding topological signal spaces \cite{vetterli2014spbook}. 

\noindent\textbf{\textit{Remark.}} From now on, we will use a simplified notation for the sake of clarity. In particular, given that there is no risk of incurring a loss of generality, we will drop the order subscript $k$ whenever there is no risk of ambiguity. For example, we will refer to a $k$-topological signal $\mathbf{y}_k \in \mathbb{R}^{N_k}$ simply as "topological signal" and denote it with $\mathbf{y} \in \mathbb{R}^{N}$.  

\noindent\textbf{Signal Translation on Cell Complexes.} Given a cell complex  $\mathcal{X}_K$ and a topological signal $\mathbf{y}$, we  rewrite the $l$-th frequency component $\hat{y}(l)$ of $\mathbf{y}$ from \eqref{eq:tft} as:
  $\hat{y}(l) = \mathbf{u}^H_l \mathbf{y}= \sum_{m=1}^N y(m) u^*_l(m),$
$l=1,\ldots,N$, where $\mathbf{u}_l$ is the $l$-th eigenvector of the $k$-th Hodge Laplacian $\mathbf{L}$. The $m$-th component of $\mathbf{y}$ is given by:
    $y(m)= \sum_{l=1}^N \hat{y}(l) u_l(m)$,
$m=1,\ldots,N$. Generalizing \cite{thanou2014learning}, we now constructively motivate the choice of using cell complex FIR filters as in \eqref{eq:sep_filter} to design overcomplete dictionaries for sparse representation. 

\noindent\textit{\textbf{Definition 7 (Generalized Translation Operator).}} Given a topological signal $\mathbf{g}$, we define the generalized translation operator $\mathbf{T}_m \in \mathbb{R}^{N \times N}$ centered at the $m$-th cell as:
\begin{equation}
\label{eq:transl}
\mathbf{T}_m\mathbf{g}=\sqrt{N}\sum_{l=1}^N \hat{g}(l)u^*_l(m)\mathbf{u}_l,
\end{equation}
$m=1,\ldots,N$, where $\hat{\mathbf{g}}=\mathbf{U}^T\mathbf{g}$ can be interpreted as a \textit{kernel} that controls the localization of the translation $\mathbf{T}_m\mathbf{g}$ around the $m$-th cell. As a consequence, we could design a smooth kernel $\hat{\mathbf{g}}$ arbitrarily localized around a cell $m$. Polynomials of the eigenvalues of the Hodge Laplacians are instrumental for this aim, as they are localized, and their order $J$  controls for their smoothness. Let us denote the set of eigenvalues of $\mathbf{L}^{(d)}$ and $\mathbf{L}^{(u)}$ with $\mathcal{F}^{(d)}$ and $\mathcal{F}^{(u)}$, respectively. Thus, following similar arguments to \eqref{eq:sep_filter}, we define the kernel as:
\begin{align}\label{eq:kernel_scalar}
\hat{g}(m) &= \bigg(\sum_{j=1}^{J} \lambda_m^j\Big(h^{(u)}_j \,\mathbb{I}(\lambda_m \in \mathcal{F}^{(u)}) \nonumber \\
    &\quad\quad+ h^{(d)}_j \,\mathbb{I}(\lambda_m \in \mathcal{F}^{(d)})\Big) + h^{(id)}\bigg),
\end{align}
$m=1,\ldots,N$, where $\mathbb{I}$ denotes the indicator function. 

\noindent\textbf{Complete Parametric Dictionaries.} The properties of localization and smoothness make each vector as in \eqref{eq:transl} with polynomial kernels as in \eqref{eq:kernel_scalar} a reasonable choice for an \textit{atom} of a dictionary \cite{thanou2014learning}. We then define a \textit{complete parametric dictionary} for topological signals as a stack of translations, one per each cell. Formally, we write in matrix form:
\begin{align}
\label{eq:poly_transl}
[&\mathbf{T}_1, \dots,\mathbf{T}_N]\mathbf{g} =\nonumber \\
&=\sqrt{N}\mathbf{U}_k\bigg(\sum_{j=1}^{J}\Big(h^{(u)}_j (\mathbf{\Lambda}_k^{(u)})^j + h^{(d)}_j (\mathbf{\Lambda}_k^{(d)})^j\Big) + h_i^{(id)}I\bigg)\mathbf{U}_k^T \nonumber\\
&=\sqrt{N}\Big( \sum_{j=1}^J \Big(h^{(d)}_{j} (\mathbf{L}^{(d)}_k)^j + h^{(u)}_{j} (\mathbf{L}^{(u)}_k)^j \Big)+h^{(id)}\mathbf{I}\Big),
\end{align}
where $\mathbf{g} = \mathbf{U}\mathbf{\hat{g}}$ and $\mathbf{\hat{g}}$ is as in \eqref{eq:kernel_scalar}. It is evident that, up to a multiplicative constant, \eqref{eq:poly_transl} represents a cell complex FIR filter, i.e., $[\mathbf{T}_1, \dots,\mathbf{T}_N]\mathbf{g} = \sqrt{N}\mathbf{H}$, with $\mathbf{H}$ as in \eqref{eq:sep_filter}. 

\noindent\textbf{Overcomplete Parametric Dictionaries.} As such, an overcomplete parametric dictionary can be defined as a concatenation of $M$ complete parametric dictionaries, i.e., of $M$ scaled cell complex FIR filters as in \eqref{eq:sep_filter}, each having a different set of filters coefficients. Formally, an overcomplete parametric dictionary for topological signals is defined as:
\begin{align}
\label{eq:dict_param}
    \mathbf{D} &=\sqrt{N}[\mathbf{H}_1,  \cdots, \mathbf{H}_M] = [\mathbf{D}_1, \cdots, \mathbf{D}_M] \in \mathbb{R}^{N \times MN}
\end{align}
where $\mathbf{H}_i$ in each subdictionary $\mathbf{D}_i = \sqrt{N}\mathbf{H}_i \in \mathbb{R}^{N \times N}$ is as in \eqref{eq:sep_filter}. We denote the kernel of the $i$-th subdictionary $\mathbf{D}_i$ with $\hat{\mathbf{g}}_i$, and we collect its  coefficients  in the vector $\mathbf{h}_i=[h_i^{(id)};\mathbf{h}_i^{(u)};\mathbf{h}_i^{(d)}] \in \mathbb{R}^{(2J+1)}$ and the coefficient of the overall dictionary $\mathbf{D}$ in the vector 
$ \mathbf{h}=[\mathbf{h}_1;...;\mathbf{h}_M] \in \mathbb{R}^{(2J+1)M}$.
The polynomial parametrization ensures the localization of the atoms in the cell domain, but it does not provide any structure for their frequency representation. Following \cite{thanou2014learning}, we then make the following sanity assumptions:

    \noindent\textbf{Assumption 1.} Each kernel is nonnegative and bounded by a constant $d > 0$ in the frequency domain, i.e.:
    \begin{equation} \label{eq:spec_const1_kernel}
    0 \leq \hat{g}_i(l) \leq d,  
    \quad  i = 1, \ldots, M, 
    \quad  l = 1, \ldots, N. 
    \end{equation}
    This is equivalent to assuming that each subdictionary is positive semi-definite and has a bounded spectral norm:
    \begin{equation}\label{eq:spec_const1}
        0 \preceq \mathbf{D}_i \preceq d\mathbf{I}, \quad i = 1, \ldots, M.
    \end{equation}
    \noindent\textbf{Assumption 2.} The set of kernels $\{\hat{\mathbf{g}}_i\}_{i=1}^M$ covers the entire frequency spectrum so that no information is lost, i.e.:
    \begin{equation}
         d - \varepsilon \leq \sum_{i=1}^{M} \hat{g}_i(l) \leq d + \varepsilon, \quad l = 1, \ldots, N, \label{eq:spec_const2_kernel}   
    \end{equation}
    where $\varepsilon$ is a small positive constant. This is equivalent to assuming that the sum of all of the sub-dictionaries $\mathbf{D}_i$ has the minimum eigenvalue lower-bounded by $d-\varepsilon$, and the maximum eigenvalue upper-bounded by $d+\varepsilon$, i.e.:
    \begin{equation}\label{eq:spec_const2}
        (d - \varepsilon)\mathbf{I} \preceq \sum_{i=1}^{M} \mathbf{D}_i \preceq (d + \varepsilon)\mathbf{I}.
    \end{equation}
These sanity assumptions on the spectral representations of the atoms are required to prove the following theoretical results and, as the reader will notice in the next section, inject prior knowledge about the
topological signals to be represented. 

\begin{lemma}\label{lemma:stability}
\textbf{(Topological frames)} Given an overcomplete dictionary $\mathbf{D}$ as in \eqref{eq:dict_param}, if the kernels $\{\hat{\mathbf{g}}_i\}_{i=1}^M$ are as in \eqref{eq:kernel_scalar} and they respect Assumptions 1 and 2, then the atoms of $\mathbf{D}$, form a frame for $\mathbb{R}^{N}$. It then holds:
\begin{equation}\label{eq:Lemma1_result}
\frac{(d-\epsilon)^2}{M}\lVert \mathbf{y} \rVert_2^2 \leq \sum_{l=1}^N \sum_{i=1}^M |\langle \mathbf{y}, [\mathbf{D}_i]^l \rangle|^2 \leq (d+\epsilon)^2 \lVert \mathbf{y} \rVert_2^2,
\end{equation}
for all topological signal $\mathbf{y} \in \mathbb{R}^N$.
\end{lemma}
\textbf{\textit{Proof.}} The proof generalizes the approach of  \cite[Proposition 1]{thanou2014learning} to our topological setup and follows similar steps. \hspace{.8cm}\qedsymbol{}



\section{Problem Formulation}
\label{sec:param_dic_learn}

In this section, we formulate the topological dictionary learning problem.
Given a training set of $T$ $k$-topological signals $\mathbf{Y} = [\mathbf{y}_1, \ldots, \mathbf{y}_T] \in \mathbb{R}^{N \times T}$, we aim to learn an overcomplete dictionary, which can represent the training signals as a sparse linear combination of the atoms. We consider an overcomplete dictionary given by the topological frames in \eqref{eq:dict_param}, which are parametrized by the vector of filters coefficients $\mathbf{h} \in \mathbb{R}^{(2J+1)M}$. In principle, a full knowledge of the cell complex structure, as defined by the upper and lower Laplacians in \eqref{eq:poly_transl}, is essential for the design of effective dictionaries, as discussed in Sec. III. However, in practical scenarios, it is often reasonable to assume access only to the lower adjacency information — that is, the graph representing the $1$-skeleton of the cell complex — whose structure is captured by the lower Laplacian $\mathbf{L}_k^{(d)}$. In contrast, higher-order information, such as the identification of active polygons among all possible induced cycles, is typically unavailable and must be learned from data in a task-specific manner. To this end, let $\mathcal{C}$ denote the set of all induced cycles in the complex (selected up to a certain order). We parametrize the upper Laplacian $\mathbf{L}_k^{(u)}$ in (\ref{Laplaciank}) using a set of indicator variables $p_j \in \{0,1\}$ for each $j \in \mathcal{C}$, which takes the value 1 if the corresponding cycle is an active polygon, and 0 otherwise. In formulas, the upper Laplacian $\mathbf{L}_k^{(u)}$ can be written as:
\begin{equation}
\label{eq:indicator_topo}
    \mathbf{L}_k^{(u)}(\mathbf{p})= \sum_{j\in \mathcal{C}} p_j \mathbf{b}_j \mathbf{b}_j^T,
\end{equation}
where $\{\mathbf{b}_j\}_{j=1}^M$ are the column vectors of the matrix $\mathbf{B}_{k+1}$ built considering all possible edge-cycle incidence relations; and $\mathbf{p}=[p_1,\ldots,p_{|\mathcal{C}|}]^T\in \{0,1\}^{|\mathcal{C}|}$ is the vector collecting all the polygon indicator variables.
Applying (\ref{eq:indicator_topo}) to (\ref{eq:dict_param}), the dictionary $\mathbf{D(h,p)}$ becomes a function of both the filter coefficients $\mathbf{h}$ and the topological structure encoded into $\mathbf{p}$.


The topological dictionary learning problem is formulated as a joint optimization w.r.t. the dictionary coefficients $\mathbf{h}$, the upper Laplacian parameters $\mathbf{p}$, and the sparse signal representation $\mathbf{S}$, aimed at minimizing a regularized data fitting term with respect to available training data $\mathbf{Y}$, while imposing the topological spectral constraints in \eqref{eq:spec_const1}-\eqref{eq:spec_const2}. Mathematically, the problem can be cast as:
\begin{align}
    \label{eq:all_opt}
    &\hspace{-.8cm}\underset{\mathbf{h,S,p}}{\arg \min} \;\;\|\mathbf{Y - D(h,p)S}\|^2_F + \gamma \|\mathbf{h}\|^2_2 \tag{$\mathbf{P}_0$}\\
    &\hspace{-.3cm}\hbox{subject to} \nonumber\\
     & (a) \;\; \|[\mathbf{S}]^i\|_0 \leq K_0, \quad i = 1, \ldots, T \nonumber\\
     & (b) \;\; 0 \preceq \mathbf{D}_i \preceq d\mathbf{I}, \quad i = 1, \ldots, M \nonumber\\ 
     & (c) \;\; (d - \varepsilon)\mathbf{I} \preceq \sum_{i=1}^{M} \mathbf{D}_i \preceq (d + \varepsilon)\mathbf{I} \nonumber\\
     & (d) \;\; \mathbf{D}_i \; \text{defined as in \eqref{eq:dict_param}}, \quad i = 1, \ldots, M \nonumber\\
     & (e) \;\; \mathbf{p} \in \{0,1\}^{|\mathcal{C}|}  \nonumber
\end{align}
where constraint (a) imposes sparsity of the signal representation $\mathbf{S}$ limiting the number of non-zero elements of each column to a maximum value $K_0$; (b) and (c) impose topological spectral constraints to ensure proper frame bounds (see Lemma 1); (d) defines the parametrization of the topological dictionary structure; finally, (e) introduces discrete constraints on the binary indicator vector $\mathbf{p}$. The data fitting term in the objective of $\mathbf{h}$ is regularized through a weighted $\ell_2$ function, with $\gamma>0$ denoting the regularization parameter, aimed at penalizing large values of the filter coefficients, mitigating overfitting, and ensuring numerical stability. 

\section{Algorithmic solutions}
Unfortunately, problem \ref{eq:all_opt} is non-convex and NP-hard, making it challenging to solve directly. To address this, we must develop an effective algorithmic strategy capable of handling the joint optimization of the three variables, $\mathbf{h}$, $\mathbf{S}$, and $\mathbf{p}$. To this end, we propose an alternating optimization approach, which decomposes the overall problem \ref{eq:all_opt} into a series of simpler sub-problems. Each sub-problem focuses on optimizing a single variable while keeping the others fixed, thereby facilitating a more tractable solution process. In the sequel, we will describe the sub-problems and the proposed alternating optimization. 


\subsection{Sub-problem 1: Dictionary coefficients}
Fixing the sparse signal representations, $\overline{\mathbf{S}}$, and the upper Laplacian parameters, $\overline{\mathbf{p}}$, problem $\mathbf{P}_0$ becomes:
\begin{align}
    \label{eq:sdp}
    &\hspace{-.8cm}\underset{\mathbf{h}}{\arg \min} \;\;\|\mathbf{Y - D(h,\overline{p})\overline{S}}\|^2_F + \gamma \|\mathbf{h}\|^2_2 \tag{$\mathbf{P}_1$}\\
    &\hspace{-.3cm}\hbox{subject to} \nonumber\\
     & (b) \;\; 0 \preceq \mathbf{D}_i \preceq d\mathbf{I}, \quad i = 1, \ldots, M \nonumber\\ 
     & (c) \;\; (d - \varepsilon)\mathbf{I} \preceq \sum_{i=1}^{M} \mathbf{D}_i \preceq (d + \varepsilon)\mathbf{I} \nonumber\\
     & (d) \;\; \mathbf{D}_i \; \text{defined as in \eqref{eq:dict_param}}, \quad i = 1, \ldots, M \nonumber
\end{align}
Problem $\mathbf{P}_1$ is a semidefinite program (SDP) that can be solved using standard convex optimization solvers \cite{boyd2004convex}. However, it is well known that the solution of SDPs might become challenging even for problems with a moderate number of variables. Thus, in the following, we recast $\mathbf{P}_1$ as a quadratic program (QP). 
        
        
        
\begin{lemma}
    \label{lemma:qp_ref}
        Problem $\mathbf{P}_1$ can be reformulated as the QP:
        \begin{align}
        \label{eq:qp}
            &\hspace{-.8cm}\underset{\mathbf{h}}{\arg \min} \;\;  \mathbf{h}^T \mathbf{Q} \mathbf{h} - \mathbf{r}^T \mathbf{h} \tag{$\mathbf{P}_{1}'$}\\
             &\hspace{-.4cm} \hbox{subject to} \nonumber\\
             &0 \leq \mathbf{I}_M \otimes \mathbf{F} \mathbf{h} \leq d \nonumber\\
             &(d - \varepsilon) \mathbf{1} \leq \mathbf{1}^T \otimes \mathbf{F}\mathbf{h} \leq (d + \varepsilon) \mathbf{1} \nonumber
        \end{align}
        with the following identifications:
        \begin{align}
            \label{eq:quad_term}
            \mathbf{Q} &= \sum_{n=1}^{N} \sum_{m=1}^{T} \mathbf{v}_{mn}\mathbf{v}^T_{mn} + \gamma \mathbf{I} \;,\\        
            \label{eq:lin_term}
            \mathbf{r} &= 2 \sum_{n=1}^{N} \sum_{m=1}^{T} [\mathbf{Y}]_{mn}\mathbf{v}_{mn}\;,
        \end{align}
     where each vector $\mathbf{v}_{mn}\in \mathbb{R}^{(2J+1)M}$ is given by 
        \begin{equation}
           \label{eq:final_aux}
          \mathbf{v}_{mn}=[\mathbf{v}^T_{1,mn}\, \mathbf{v}^T_{2,mn}\, \ldots \, \mathbf{v}^T_{M,mn}]^T \in \mathbb{R}^{(2J+1)M}
        \end{equation}
        where
        \begin{align}
        \label{eq:aux_P}
         &\mathbf{v}_{i,mn} = \left[[\mathbf{I}]_m[\overline{\mathbf{S}}_i]^n\, (\mathbf{v}_{{i,mn}}^{(u)})^T \,(\mathbf{v}_{{i,mn}}^{(d)})^T \right]^T\in \mathbb{R}^{J} \\
         &\mathbf{v}_{{i,mn}}^{(u)}= \left[ [\mathbf{L}_k^{(u)}(\overline{\mathbf{p}})]_m [\overline{\mathbf{S}}_i]^n\, \ldots\, [(\mathbf{L}_k^{(u)}(\overline{\mathbf{p}}))^J]_m [\overline{\mathbf{S}}_i]^n \right]^T \in \mathbb{R}^{J}\nonumber\\
         &\mathbf{v}_{{i,mn}}^{(d)}= \left[ [\mathbf{L}_k^{(d)}]_m [\overline{\mathbf{S}}_i]^n\, \ldots\, [(\mathbf{L}_k^{(d)})^J]_m [\overline{\mathbf{S}}_i]^n \right]^T\in \mathbb{R}^{2J+1} \nonumber      
        \end{align}
        with $\overline{\mathbf{S}}_i\in \mathbb{R}^{N \times T}$ denoting the set of rows of $\overline{\mathbf{S}}$ corresponding to the atoms in the $i$-th sub-dictionary $\mathbf{D}_i$. Finally, matrix $\mathbf{F}$ for the spectral constraints in $\mathbf{P}_1'$ reads as:
        \begin{align}
        \mathbf{F} &= \left[\mathbf{1}; \mathbf{F}^{(u)}; \mathbf{F}^{(d)}\right] \in \mathbb{R}^{M \times (2J+1)} \;,\label{eq:vand_constr1}\\
        \mathbf{F}^{(u)} &= \begin{bmatrix}
        \lambda_1^{(u)} & (\lambda_1^{(u)})^2 & \ldots & (\lambda_1^{(u)})^J \\
        \lambda_2^{(u)} & (\lambda_2^{(u)})^2 & \ldots & (\lambda_2^{(u)})^J \\
        \vdots & \vdots & \ddots & \vdots \\
        \lambda_N^{(u)} & (\lambda_N^{(u)})^2 & \ldots & (\lambda_N^{(u)})^J \\
        \end{bmatrix}, \label{eq:vand_constr2} \\
        \mathbf{F}^{(d)} &= \begin{bmatrix}
        \lambda_1^{(d)} & (\lambda_1^{(d)})^2 & \ldots & (\lambda_1^{(d)})^J \\
        \lambda_2^{(d)} & (\lambda_2^{(d)})^2 & \ldots & (\lambda_2^{(d)})^J \\
        \vdots & \vdots & \ddots & \vdots \\
        \lambda_N^{(d)} & (\lambda_N^{(d)})^2 & \ldots & (\lambda_N^{(d)})^J \\
        \end{bmatrix}. \label{eq:vand_constr3}
            \end{align}
where $\lambda_l^{(u)}=\lambda_l \,\mathbb{I}(\lambda_l \in \mathcal{F}^{(u)})$ and $\lambda_l^{(d)}=\lambda_l \,\mathbb{I}(\lambda_l \in \mathcal{F}^{(d)})$ for all $l=1,\ldots,N$.
\end{lemma}

\vspace{.1cm}
\textbf{\textit{Proof.}} See Appendix \ref{appendixB}. \hspace{4.5cm}\qedsymbol{}

\vspace{.1cm}
\subsection{Sub-problem 2: Sparse coding} In this case, the sparse representation $\mathbf{S}$ is updated by solving problem \ref{eq:all_opt}, while keeping the dictionary coefficients $\overline{\mathbf{h}}$ and the (upper) topological structure $\overline{\mathbf{p}}$ fixed. Specifically, letting $\overline{\mathbf{D}}=\mathbf{D}(\overline{\mathbf{h}},\overline{\mathbf{p}})$, we obtain:
\begin{align}\label{eq:omp}
    &\hspace{-.8cm}\underset{\mathbf{S}}{\arg \min} \;\;\|\mathbf{Y - \overline{D}S}\|^2_F \tag{$\mathbf{P}_2$}\\
    &\hspace{-.3cm} \hbox{subject to} \nonumber\\
     & (a) \; \|[\mathbf{S}]^i\|_0 \leq K_0, \quad i = 1, \ldots, T \nonumber
\end{align}
\ref{eq:omp} is a \textit{sparse coding} sub-problem that is typically used in compressive sensing to find the best $K_0$-sparse approximation of signals \cite{foucartCS}. Despite being NP-hard, \ref{eq:omp} can be addressed efficiently using greedy algorithms, with \textit{Orthogonal Matching Pursuit} (OMP) \cite{cai2011orthogonal} being a widely adopted approach. 

\textit{Remark:} A key step in the OMP algorithm is dictionary normalization, which ensures a fair selection process among the atoms \cite{Elad2010}. Specifically, we address problem \ref{eq:omp} using OMP with an input dictionary $\mathbf{D}_w = \overline{\mathbf{D}} \mathbf{W}$, where $\mathbf{W}$ is a diagonal matrix whose elements are the reciprocals of the norms of the columns of $\overline{\mathbf{D}}$. This normalization adjusts the influence of each atom during the selection process. The OMP output, denoted as $\mathbf{S}_w$, is then premultiplied by $\mathbf{W}$ to obtain the final sparse representation $\mathbf{S}=\mathbf{W}\mathbf{S}_w$. This approach preserves the sparsity pattern of $\mathbf{S}_w$ in $\mathbf{S}$, and ensures that $\mathbf{D}_w\mathbf{S}_w = \overline{\mathbf{D}}\mathbf{S}$.


\subsection{Sub-problem 3: Topology learning} Here, we optimize $\mathbf{P}_0$ with respect to the upper Laplacian parameters $\mathbf{p}$. Consistent with previous steps, the other two optimization variables, $\overline{\mathbf{h}}$ and $\overline{\mathbf{S}}$, are kept fixed. The resulting sub-problem reads as:
\begin{align}
\label{eq:ip}
    &\hspace{-.8cm}
    \underset{\mathbf{p}\in \{0,1\}^{|\mathcal{C}|}}{\arg \min}\;\; \|\mathbf{Y - D(\overline{h},p)\overline{S}}\|^2_F \tag{$\mathbf{P}_3$}
\end{align}
Problem \ref{eq:ip} constitutes a non-linear integer programming problem and is NP-hard. To address this computational challenge, we propose two distinct strategies. The first strategy employs a greedy approach to iteratively select the active polygons within the cell complex, while the second strategy addresses Problem \ref{eq:ip} by solving its continuous relaxation. In both cases, we have to carefully regulate the update frequency of the vector $\mathbf{p}$ relative to the other variables, $\mathbf{h}$ and $\mathbf{S}$, to prevent potential issues with domain adaptation. The detailed implementation of the proposed algorithms is presented in the following paragraphs.

\noindent \textbf{Algorithm 1 : Greedy Topology Inference.} The rationale behind this first algorithm is simple: starting from the full topology, we iteratively remove polygons in a greedy manner while simultaneously optimizing the dictionary coefficients and sparse representations. The complete procedure is outlined in Algorithm \ref{algo:JTDL}, referred to as the Greedy Topological Dictionary Learning (GTDL) algorithm. Specifically, the process begins with a fully active topology, represented by $\mathbf{p} = \mathbf{1}$ (Lines 2-4). With $\mathbf{p}$ held fixed, the algorithm alternates between solving \ref{eq:qp} and \ref{eq:omp} until a convergence criterion is met (Lines 8-11).  
Once the solutions to \ref{eq:qp} and \ref{eq:omp} are obtained, the algorithm identifies the polygon whose removal yields the largest reduction of a meaningful function $f(\cdot)$ (e.g., the objective function of $\mathbf{P}_0$, or the mean-square error over an available test dataset), and this polygon is subsequently deactivated (lines 15-17). The process is repeated until no further benefit is obtained from removing a polygon or until the upper topology becomes empty, i.e., $\mathbf{p} = \mathbf{0}$. While this topology learning strategy has been empirically validated as robust across various sparsity levels in both the topological and signal domains (cf. Sec. VI), its computational cost can be substantial, particularly for high-dimensional cell complexes. A complexity analysis of each subproblem is detailed in the sequel:
\begin{itemize}
    \item The resolution of the QP problem in \ref{eq:qp} using any interior-point solver requires $\mathcal{O}\left(\sqrt{m}\,n^3\log\left(\frac{1}{\epsilon}\right)\right)$ \cite{ben2001lectures}, where $n=(2J+1)M$ denotes the dimensionality of the optimization variable $\mathbf{h}$, $m=\mathcal{O}(M)$ represents the number of constraints, and $\epsilon$ is the tolerance parameter that determines when the algorithm has converged sufficiently close to the optimum. Consequently, the complexity of the QP step is $\mathcal{O}\left(M^{3/2}J^3\log\left(\frac{1}{\epsilon}\right)\right)$.
    \item For the OMP step defined to handle \ref{eq:omp}, a dictionary consisting of $MN$ atoms is constructed, each representing a $k$-topological signal of dimension $N$ (i.e., defined over $N$ $k$-cells of a cell complex). The naive implementation of OMP requires a computational complexity of $\mathcal{O}\Bigl(N(MN) + (MN)K_0 + NK_0^2 + K_0^3\Bigr)$ \cite{sturm2012comparison}.
    \item In the greedy step, with $\mathbf{h}$ and $\mathbf{S}$ fixed, the primary computational task is to evaluate a meaningful function such as $f(\mathbf{h},\mathbf{S},\mathbf{p})=\|\mathbf{X}-\mathbf{D}(\mathbf{h},\mathbf{p})\,\mathbf{S}\|_2^2$ for different candidate modifications of $\mathbf{p}$ at each iteration. It can be shown that the cost of computing $f(\cdot)$ is $\mathcal{O}\left(|\mathcal{C}|N^2+JMN^3\right)$, where $|\mathcal{C}|$ denotes the number of $(k+1)$-cells (in our numerical experiments, $\mathcal{C}$ represents the set of polygons, i.e., 2-cells). Moreover, since one active component is removed from $\mathbf{p}$ in each iteration, the number of candidate modifications of $\mathbf{p}$ to be evaluated at the $i$-th step is $|\mathcal{C}|-i$. In the worst case, the algorithm performs $|\mathcal{C}|$ iterations, resulting in a total of $\sum_{i=0}^{|\mathcal{C}|} \Bigl(|\mathcal{C}|-i\Bigr)=\mathcal{O}\left(|\mathcal{C}|^2\right)$ evaluations. Thus, the overall computational cost of topology inference via greedy search is $\mathcal{O}\left(|\mathcal{C}|^3N^2+|\mathcal{C}|^2JMN^3\right)$.
\end{itemize}
Taking into account that the dimensional hyperparameters $J$ and $M$ and the assumed sparsity $K_0$ are typically much smaller than the topological dimensions $|\mathcal{C}|$ and $N$, Algorithm \ref{algo:JTDL} exhibits an overall complexity of $\mathcal{O}\left(|\mathcal{C}|N^2I_{\text{max}}+|\mathcal{C}|^3N^2+|\mathcal{C}|^2N^3\right)$, scaling cubically both with the number of $(k+1)$-cells and the number of $k$-cells.

\begin{algorithm}[t]
\caption{: Greedy Topological Dictionary Learning}
\label{algo:JTDL}
\begin{algorithmic}[1]
\Statex \textbf{Inputs:}
\Statex \quad $\mathbf{Y} \in \mathbb{R}^{M \times N}$: Training signals.
\Statex \quad $I_{\text{max}}$: Number of iterations (stopping criterion).
\Statex \quad $K_0$: Assumed sparsity level.
\Statex \textbf{Outputs:}
\Statex \quad $\mathbf{h}^*$: Dictionary coefficients.
\Statex \quad $\mathbf{S}^*$: Sparse representations.
\Statex \quad $\mathbf{p}^*$: Upper Laplacian parameters.
\Procedure{TDL}{\textbf{Inputs}}
\State Initialize $\mathcal{P} \leftarrow \{1, \cdots, |\mathcal{C}|\}$, and set $\mathbf{p}_{\mathcal{P}}=\mathbf{1}_{\mathcal{P}}$
\State Initialize $err,err' \leftarrow \infty$
\State Initialize $\overline{\mathbf{h}}$, $\overline{\mathbf{S}}$ at random.
\While{$\mathcal{P} \neq \emptyset$ and $err \geq err'$}\smallskip
    \State Evaluate $\mathbf{L}_k^{(u)}$ from $\mathcal{P}$ as in \eqref{eq:indicator_topo}\smallskip
    \State Set $\mathbf{h}^{[0]}=\overline{\mathbf{h}}$ and $\mathbf{S}^{[0]}=\overline{\mathbf{S}}$
    \For{$t = 0$ \textbf{to} $I_{\text{max}}$}
    \State Get $\mathbf{h}^{[t+1]}$ from solving \ref{eq:qp} 
    \State Get $\mathbf{S}^{[t+1]}$ from solving \ref{eq:omp}
\EndFor
    \State Set $\mathbf{\overline{h}}= \mathbf{h}^{[I_{\text{max}}]}$ and $\mathbf{\overline{S}}= \mathbf{S}^{[I_{\text{max}}]}$
    \State $err \leftarrow f(\mathbf{\overline{h}},\overline{\mathbf{S}},\mathbf{p}_{\mathcal{P}})$\smallskip
    \State \textbf{\textit{Topology Update} (Greedy Search):}\smallskip
    \State $p^* \leftarrow \underset{j \in \mathcal{P}}{\arg\min} \;f(\mathbf{\overline{h}},\overline{\mathbf{S}},\mathbf{p}_{\mathcal{P}\setminus\{j\}})$
    \State $err' \leftarrow f(\mathbf{\overline{h}},\overline{\mathbf{S}},\mathbf{p}_{\mathcal{P}\setminus\{p^*\}})$
    \State $\mathcal{P} \leftarrow \mathcal{P} \setminus \{p^*\}$ and $\mathbf{p}_{\mathcal{P}}=\mathbf{1}_{\mathcal{P}}$
\EndWhile
\Statex \textbf{return}
\State $ \quad\mathbf{h}^* = \overline{\mathbf{h}}$
\State $ \quad \mathbf{S}^* = \overline{\mathbf{S}}$
\State $ \quad \mathbf{p}^* = \mathbf{p}_{\mathcal{P}}$
\EndProcedure
\end{algorithmic}
\end{algorithm}

 \begin{algorithm}[t]
\caption{: Relaxed Topological Dictionary Learning}
\label{algo:STDL}
\begin{algorithmic}[1]
\Statex \textbf{Inputs:}
\Statex \quad $\mathbf{Y} \in \mathbb{R}^{M \times N}$: Training signals.
\Statex \quad $I_{\text{max}}$: Number of iterations (stopping criterion).
\Statex \quad $K_0$: Assumed sparsity level.
\Statex \textbf{Outputs:}
\Statex \quad $\mathbf{h}^*$: Dictionary coefficients.
\Statex \quad $\mathbf{S}^*$: Sparse representations.
\Statex \quad $\mathbf{p}^*$: Upper Laplacian parameters.
\Procedure{TDL}{\textbf{Inputs}}
\State Initialize $\mathbf{p}^{[0]}=\underline{\mathbf{1}}$
\State Initialize $\mathbf{h}^{[0]}$, $\mathbf{S}^{[0]}$ at random.
\For{$t = 0$ \textbf{to} $I_{\text{max}}$}
    \State Get $\mathbf{h}^{[t+1]}$ from solving \ref{eq:qp}
    \State Get $\mathbf{S}^{[t+1]}$ from solving \ref{eq:omp}
    \State $\mathbf{p}^{[t+1]} \leftarrow \mathcal{H}^{[0,1]}_{\lambda}\left(\mathbf{p}^{[t]} - \mu \nabla f\left(\mathbf{h}^{[t+1]},\mathbf{S}^{[t+1]},\mathbf{p}^{[t]}\right)\right)$ 
    \State Evaluate $\mathbf{L}_k^{(u)}$ from $\mathbf{p}^{[t+1]}$ as in \eqref{eq:indicator_topo}\smallskip
\EndFor
\Statex \textbf{return}
\State $ \quad\mathbf{h}^* = \mathbf{h}^{[I_{\text{max}}]}$
\State $ \quad \mathbf{S}^* = \mathbf{S}^{[I_{\text{max}}]}$
\State $ \quad \mathbf{p}^* = \mathbf{p}^{[I_{\text{max}}]}$
\EndProcedure
\end{algorithmic}
\end{algorithm}

\noindent \textbf{Algorithm 2 : Relaxed Topology Inference.} The greedy selection in Algorithm 1 for solving problem \ref{eq:ip} can be computationally expensive. To mitigate this, we relax the discrete constraint in \ref{eq:ip}, reformulating it as a continuous optimization problem with respect to the polygon variable vector $\mathbf{p}$, which we also assume to be sparse.
Thus, we obtain the following relaxed problem:
\begin{align}
\label{eq:relaxed_ip}
    &\hspace{-.8cm}
    \underset{\mathbf{p}\in [0,1]^{|\mathcal{C}|}}{\arg \min}\;\; \|\mathbf{Y - D(\overline{h},p)\overline{S}}\|^2_F + \lambda G(\mathbf{p}) \tag{$\mathbf{P}_3'$}
\end{align}
where $G$ is a sparsity-inducing penalty (SIP) function, which can be selected among many possible candidates \cite{foucartCS}. We then proceed minimizing $\mathbf{P}_3'$ by means of a proximal gradient descent algorithm, whose proximal operator is designed to take into account the SIP $G$ and the box constraint. As an example, considering a hard thresholding proximal operator boxed into the interval $[0,1]$, we end up into the following parameter update:
\begin{equation}\label{eq:prox_grad}
    \mathbf{p}^{[t+1]}=\mathcal{H}^{[0,1]}_{\lambda}\left(\mathbf{p}^{[t]} - \mu \nabla f\left(\mathbf{h}^{[t+1]},\mathbf{S}^{[t+1]},\mathbf{p}^{[t]}\right)\right)
\end{equation}
where $\mu>0$ is a constant step-size, and $\mathcal{H}^{[0,1]}_{\lambda}(\cdot)$ reads as:
\begin{align}\label{eq:prox_op}
\mathcal{H}^{[0,1]}_{\lambda}(\mathbf{z}) = \operatorname{Prox}^{[0,1]}_{\lambda G}(\mathbf{z}) &=
\begin{cases}
1, &  \text{if } z_{i}\geq 1\\
z_{i}, & \text{if } \sqrt{2\lambda}\leq z_{i}< 1, \\
0, & \text{if }  z_{i}< \sqrt{2\lambda},
\end{cases} 
\end{align}
with $0<\sqrt{2\lambda}<1$. Other specific designs are clearly possible, but we found (\ref{eq:prox_op}) very effective in our numerical experiments. The overall procedure is detailed in \ref{algo:STDL}, referred to as the Relaxed Topological Dictionary Learning (RTDL) algorithm.
The proposed approach allows to efficiently address the complexity and domain adaptation issues inherent to the greedy version. In particular, it enables uniform frequency updates for the dictionary coefficients (line 5), sparse representations (line 6), and the upper Laplacian (lines 7-9), while controlling the optimization speed for $\mathbf{p}$ via the step size parameter $\mu$. In terms of topology update, Algorithm \ref{algo:STDL} exhibits a computational complexity of $\mathcal{O}(|\mathcal{C}|N^3I_{\text{max}})$, primarily attributable to the gradient evaluation for the update of variable $\mathbf{p}$ executed at line 7. The principal advantage of this second TDL methodology is that it guarantees linear scaling w.r.t. $|\mathcal{C}|$.

\section{Numerical Results}

This section evaluates the proposed method's performance on synthetic and real data, focusing on second-order cell complexes for efficient implementation and visualization while allowing generalization to higher-order domains. We analyze the accuracy-compression trade-off across dictionary parameterizations, assess algorithm convergence, demonstrate topological structure learning, and apply the method to real-world data, showcasing its advantages over existing approaches. The numerical results were obtained using our publicly available online code to ensure reproducibility\footnote{Code available at \href{https://github.com/SPAICOM/topological-dictionary-learning}{github.com/SPAICOM/topological-dictionary-learning}}.

\subsection{Accuracy-compression trade-off}

In this paragraph, we assess the reconstruction performance of our dictionary learning procedure relative to the number of parameters used to represent the data. 

We construct multiple second-order cell complexes to generate synthetic data. First, we build a graph with 40 vertices and 100 edges using the Erdős-Rényi model. From this graph, we probabilistically include 2-cells, defined by induced cycles, with a probability $q_{tr}$, restricting selection to triangular cycles for simplicity. Varying this probability produces different cell complexes that share the same 1-skeleton but differ in the number of triangles. For each resulting cell complex, we generate $10$ datasets, each containing $T=220$ edge signals with fixed sparsity, represented as $\mathbf{Y} = [\mathbf{y}_1, \ldots, \mathbf{y}_T] \in \mathbf{R}^{N \times T}$, where $N=100$ is the number of edges, and the generating function is represented by a second-order Hodge Laplacian polynomial of the corresponding topology. We then divide each dataset into $T_{tr}=150$ training signals and $T_{te}=80$ test signals. Specifically, for each dataset, we generate the Hodge Laplacians and their powers, i.e., $\{(\mathbf{L}^{(d)}_k)^1, (\mathbf{L}^{(u)}_k)^1, \dots, (\mathbf{L}^{(d)}_k)^J, (\mathbf{L}^{(u)}_k)^J\}$. Additionally, we store the corresponding maximum and minimum eigenvalues:
\begin{align*}
\boldsymbol{\lambda}_{\max}^{(d)}=[\lambda_{\max}^{(d)} \dots (\lambda_{\max}^{(d)})^J]^T, \;\; \boldsymbol{\lambda}_{\min}^{(d)}=[\lambda_{\min}^{(d)} \dots (\lambda_{\min}^{(d)})^J]^T, \\
\boldsymbol{\lambda}_{\max}^{(u)}=[\lambda_{\max}^{(u)} \dots (\lambda_{\max}^{(u)})^J]^T, \;\; \boldsymbol{\lambda}_{\min}^{(u)}=[\lambda_{\min}^{(u)} \dots (\lambda_{\min}^{(u)})^J]^T.
\end{align*} 
We randomly generate the parameters $\mathbf{h}^{(d)}, \mathbf{h}^{(u)}, h^{(id)}$ for $M$ different sub-dictionaries (cf. Sec. II), normalizing them by the corresponding maximum eigenvalue in $\Lambda_{\max}^{(d)}$ and $\Lambda_{\max}^{(u)}$, respectively. This completes the generation of the topological dictionaries $\mathbf{D}(\mathbf{h},\mathbf{t})$. Now,  we randomly generate also the sparse representation $\mathbf{S}$, with each column $[\mathbf{S}]^i$ presenting a fixed number of non-zero entries. Finally, we get the topological signals by applying the model as $\mathbf{Y} =\mathbf{D}(\mathbf{h},\mathbf{t})\mathbf{S}$.

To solve problem \ref{eq:all_opt} from the given dictionary structure and the data $\mathbf{Y}$, we set the control parameters $d$ and $\varepsilon$ as follows. 
Letting $\mathbf{H}^{(d)} = [\mathbf{h}^{(d)}_1, \dots, \mathbf{h}^{(d)}_M] \in \mathbb{R}^{M \times J}$, $\mathbf{H}^{(u)} = [\mathbf{h}^{(u)}_1, \dots, \mathbf{h}^{(u)}_M] \in \mathbb{R}^{M \times J}$, and $\mathbf{h}^{id} = [h^{(id)}_1 \dots h^{(id)}_M] \in \mathbb{R}^{M}$, we introduce the vectors:
\begin{align*}
    \mathbf{g_{\max}} &= \mathbf{H}^{(d)} \boldsymbol{\lambda}_{\max}^{(d)} + \mathbf{H}^{(u)} \boldsymbol{\lambda}_{\max}^{(u)} + \mathbf{h}^{id} \\
    \mathbf{g_{\min}} &= \mathbf{H}^{(d)}, \boldsymbol{\lambda}_{\min}^{(d)} + \mathbf{H}^{(u)} \boldsymbol{\lambda}_{\min}^{(u)} + \mathbf{h}^{id},
\end{align*}
and set $d$ as the maximum entry value of $\mathbf{g_{\max}}$; whereas, $\varepsilon$ is defined as $\varepsilon = \min \{ \Delta_{\min}, \Delta_{\max} \}$, where:
\begin{equation}
    \Delta_{\max} = d - \sum_{i=1}^M [\mathbf{g_{\max}}]_i, \;\;
    \Delta_{\min} = \sum_{i=1}^M [\mathbf{g_{\min}}]_i - d.
\end{equation}
In our case, we set $J=2$ and $M=3$.

\begin{figure}[t]
\vspace{-.7cm}
    \centering
    \includegraphics[width=1.\linewidth]{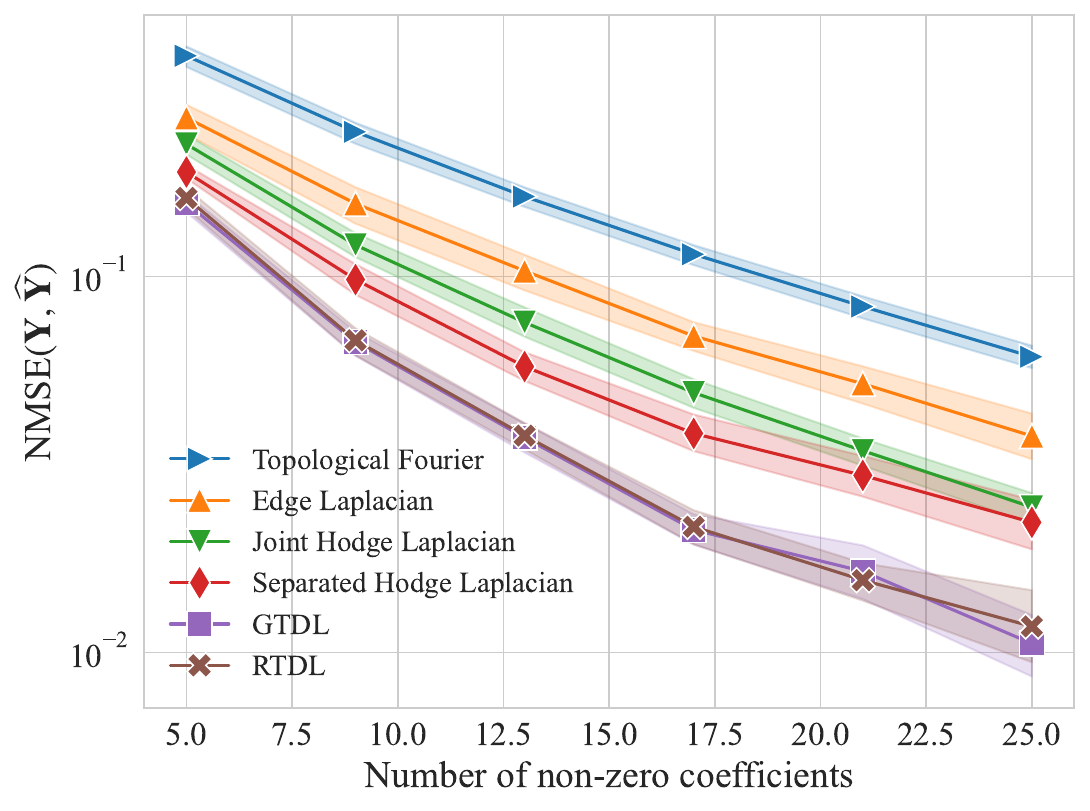}
    \caption{NMSE versus number of non-zero coefficients, for different sparse representation strategies. 
    }
    \label{fig:SyntheticSignals}
\end{figure}

In the sequel, we assess the sparse representation capabilities of our algorithms for several sparsity levels, highlighting the trade-off between accuracy and signal compressibility. Figure \ref{fig:SyntheticSignals} presents the signal reconstruction error for various sparse representation techniques, averaged over 10 datasets of signals. These signals are generated with a fixed number of non-zero coefficients (i.e., 25) and are defined over the edges of a topology containing 70\% of the possible triangles. We use a regularization term of $\gamma = 10^{-7}$ for all the dictionary learning methods, and a $\lambda=0.045$ for the proximal operator of the RTDL procedure. Given the signal reconstructions of the test set $\mathbf{\widehat{Y}} = [\mathbf{\widehat{y}}_1, \ldots, \mathbf{\widehat{y}}_N] \in \mathbb{R}^{M \times T_{te}}$, the sparse approximation error is evaluated using the Normalized Mean-Squared Error (NMSE), defined as:
\begin{equation}
\label{eq:Y_NMSE}
\operatorname{NMSE}(\mathbf{Y, \widehat{Y}})=\frac{1}{T_{te}} \sum_{n=1}^{T_{te}} \frac{\|\mathbf{y}_{n}- \mathbf{\hat{y}}_{n}\|^2_{2}}{\|\mathbf{y}_{n}\|^2_{2}}.
\end{equation}
We evaluate our approach using different dictionary parameterization techniques: (i) the simple topological Fourier Transform \cite{schaub2020random}; (ii) the Edge Laplacian, which constructs the dictionary from atoms as in (\ref{eq:sep_filter}), ignoring upper topological information (i.e., using only $(\mathbf{L}^{(d)})$; (iii) the joint Hodge Laplacian, which builds the dictionary from filters as in (\ref{eq:joint_filter}), without separating upper and lower components; (iv) the separated Hodge Laplacian strategy, which constructs the dictionary from atoms as in (\ref{eq:sep_filter}); and (v) the proposed GTDL and RTDL strategies outlined in Algorithms 1 and 2. All the methods that do not infer the topological domain assume all triangles are present in the upper Laplacian.  

As observed in Fig. \ref{fig:SyntheticSignals}, learnable dictionaries achieve a significant improvement over the simple topological Fourier Transform \cite{schaub2020random}. By comparing different dictionary parameterization techniques, Fig. \ref{fig:SyntheticSignals} empirically demonstrates that:
(i) Lower adjacency information alone has limited representational capability, whereas the proposed dictionary parameterization using higher-order cell FIR filters leads to better signal reconstruction; (ii) Separately parameterizing the upper and lower Laplacians, as in Eq. (\ref{eq:sep_filter}), provides greater flexibility to the algorithm, improving its performance; (iii) Joint learning of the sparse representation and the upper Laplacian structure enhances performance, showing that integrating topology and dictionary learning surpasses conventional topological dictionary learning, even when using the same Separated Hodge Laplacian parameterization; (iv) The proposed approaches defined in Algorithms \ref{algo:JTDL} and \ref{algo:STDL}, respectively, yield comparable performance in terms of signal approximation.


\subsection{Convergence behavior}

\begin{figure}
    \centering
    \includegraphics[width=1.\linewidth]{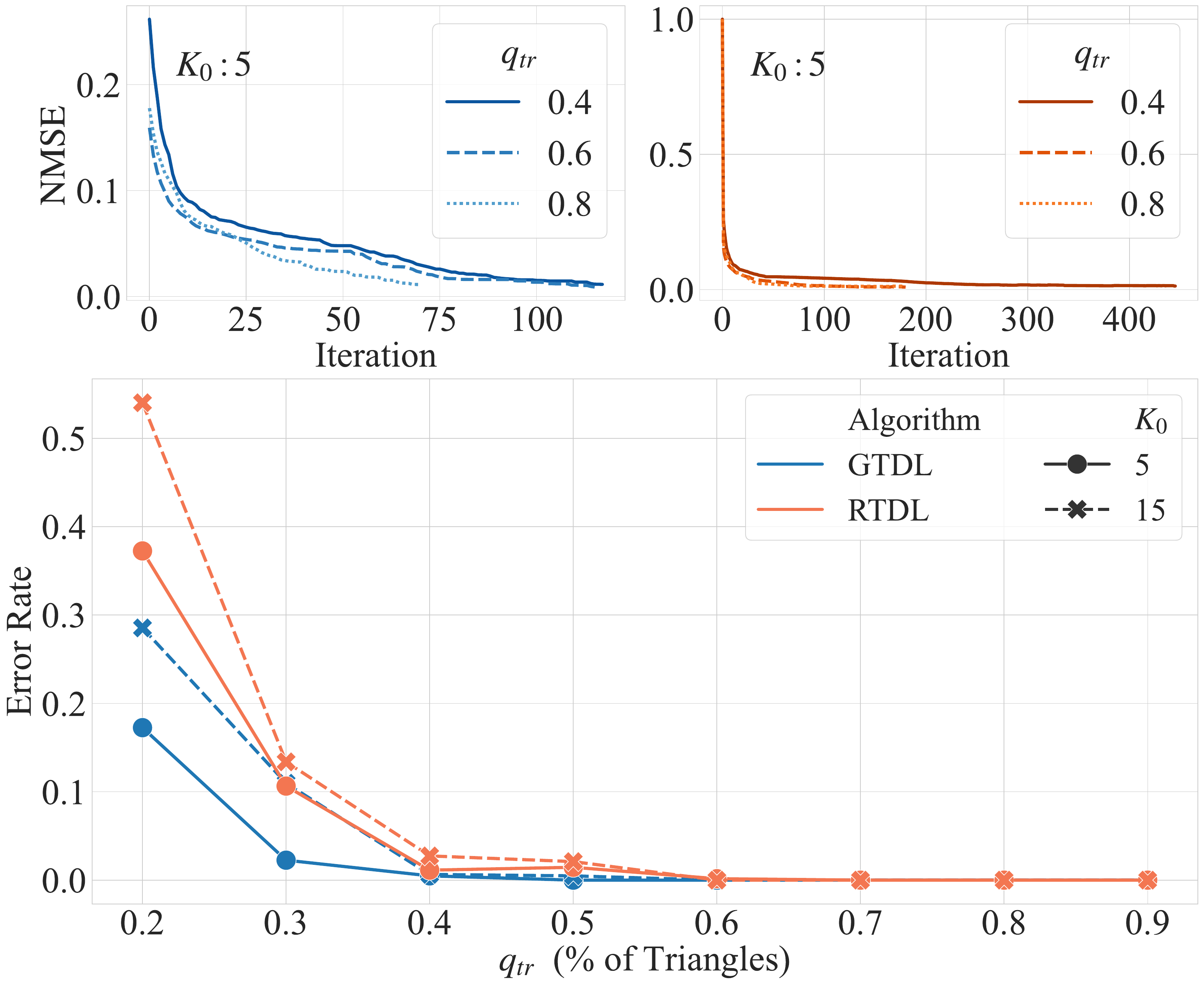}
    \caption{(Top) Objective versus iteration index, for different algorithms and upper topological structure. (Bottom) Topology approximation error versus percentage of triangles in the true topology, for different algorithms and sparsity levels.}
    \label{fig:TopoLearn}
\end{figure}

In this paragraph, we analyze the convergence properties of our dictionary learning procedures. The proposed algorithms demonstrate favorable convergence behavior, driven by the decreasing trend of the objective function in $\mathbf{P}_0$ over time. Specifically, every step of the GTDL and RTDL procedures ensures a reduction in the objective value at each iteration. The dictionary coefficient update involves solving the strongly convex problem  $\mathbf{P}'_1$, achieving the minimum objective value while keeping other variables fixed. For GTDL, the topology update is explicitly designed to reduce the objective value at each iteration. Similarly, in RTDL, selecting a sufficiently small step size $\mu$ (smaller than $2/L_t$, where $L_t$ is the Lipschitz constant of the objective with respect to $\mathbf{t}$) in (\ref{eq:prox_grad}) guarantees a decrease in the objective value at each iteration. Finally, the sparse coding step in $\mathbf{P}_2$ is updated using OMP, with the update applied only if it results in a reduction of the objective value; otherwise, the previous $K_0$-sparse solution is retained. Since the objective of $\mathbf{P}_0$ is coercive and therefore bounded from below, the sequence of algorithm iterates is guaranteed to converge. An example of the convergence behavior of the proposed GTDL and RTDL methods is shown in Fig. \ref{fig:TopoLearn} (top), considering a sparsity value of $K_0=5$ and varying percentages $q_{tr}$ of triangles present in the true topology. As observed in Fig. \ref{fig:TopoLearn} (top), both methods exhibit a smooth and well-controlled convergence in all tested conditions.

\begin{figure*}
    \centering
    \includegraphics[width=.93\linewidth]{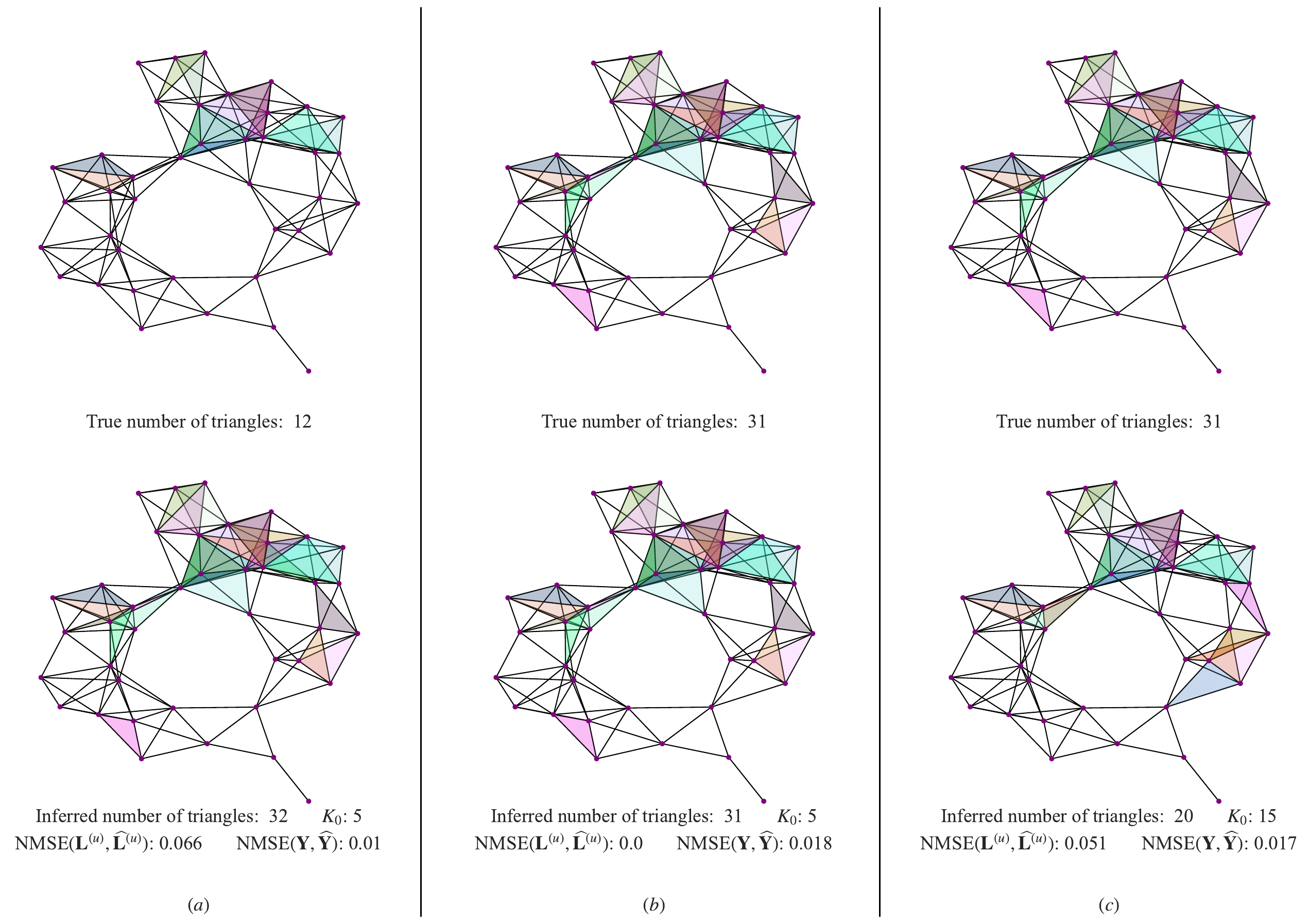}
    \caption{Visualization of topology inference performance using the GTDL algorithm, considering three distinct learning scenarios. 
    }
    \label{fig:TopoLearnBig}
\end{figure*}


\subsection{Topology learning}
In this section, we assess the performance of our algorithms to learn the underlying topological domain from data. To this aim, let us introduce a metric of topology estimation performance in terms of error rate:
\begin{equation}
\label{eq:error_rate}
    \operatorname{Error Rate}:= \frac{R}{|\mathcal{C}|},
\end{equation}
where $R$ represents the number of erroneously inferred polygons, and $|\mathcal{C}|$ is the total number of polygons in the complex. Then, in Fig. \ref{fig:TopoLearn} (bottom), we illustrate the topology error rate defined in (\ref{eq:error_rate}), obtained using the proposed GTDL and RTDL methods, as a function of the percentage of triangles in the true underlying topology. The results are averaged over 10 different realizations, and are shown for different sparsity values in the data generation process, specifically $K_0=5$ and $K_0=15$. As observed in Fig. \ref{fig:TopoLearn} (bottom), both GTDL and RTDL methods can perfectly infer the true upper topology when the percentage of triangles is sufficiently high. Conversely, as the percentage of triangles in the true topology decreases, the methods begin to introduce errors. Notably, in this more challenging scenario, GTDL outperforms RTDL, albeit at the cost of increased computational complexity. Furthermore, as seen in Fig. \ref{fig:TopoLearn} (bottom), lower sparsity levels in the true underlying data facilitate topology inference. Interestingly, while a higher number of nonzero coefficients reduces the signal approximation error, it negatively impacts topology inference, exhibiting an inverse trend. This can be explained by the fact that increasing the degrees of freedom in signal representation (i.e., higher values of $K_0$) allows multiple topologies to achieve the same reconstruction error more easily, thereby introducing ambiguity in the inference process. 

A pictorial example of the results of topology inference with GTDL is shown in Fig. \ref{fig:TopoLearnBig}, illustrating three different scenarios: (a) The topology contains 20\% of triangles, and $K_0=5$; (b) The topology contains 50\% of triangles, and $K_0=5$; (c) The topology contains 20\% of triangles, but $K_0=15$. The true topologies are depicted in the top row, whereas the inferred ones are shown in the bottom row. The topology inference performance are also expressed in terms of upper Laplacian approximation error $\operatorname{NMSE}(\mathbf{L}^{(u)}_{k}, \mathbf{\widehat{L}}^{(u)}_{k})= \frac{\|\mathbf{L}^{(u)}_{k}- \mathbf{\widehat{L}}^{(u)}_{k}\|_{F}}{\|\mathbf{L}^{(u)}_{k}\|_{F}},$
where $\mathbf{\widehat{L}}^{(u)}_{k}$ is the upper Laplacian estimated by the proposed algorithms. Similar to the observations in Fig. \ref{fig:TopoLearn} (bottom), when a sufficiently large number of true triangles is present and $K_0$ is low (see case (b) in Fig. \ref{fig:TopoLearnBig}), the topology is perfectly recovered. Conversely, when the number of true triangles is reduced (see case (a) in Fig. \ref{fig:TopoLearnBig}), or when $K_0$ is increased (see case (c) in Fig. \ref{fig:TopoLearnBig}), the topology might be either under- or over-estimated.

\subsection{Application to real data}
\begin{figure}[t]
    \centering
    \includegraphics[width=1.\linewidth]{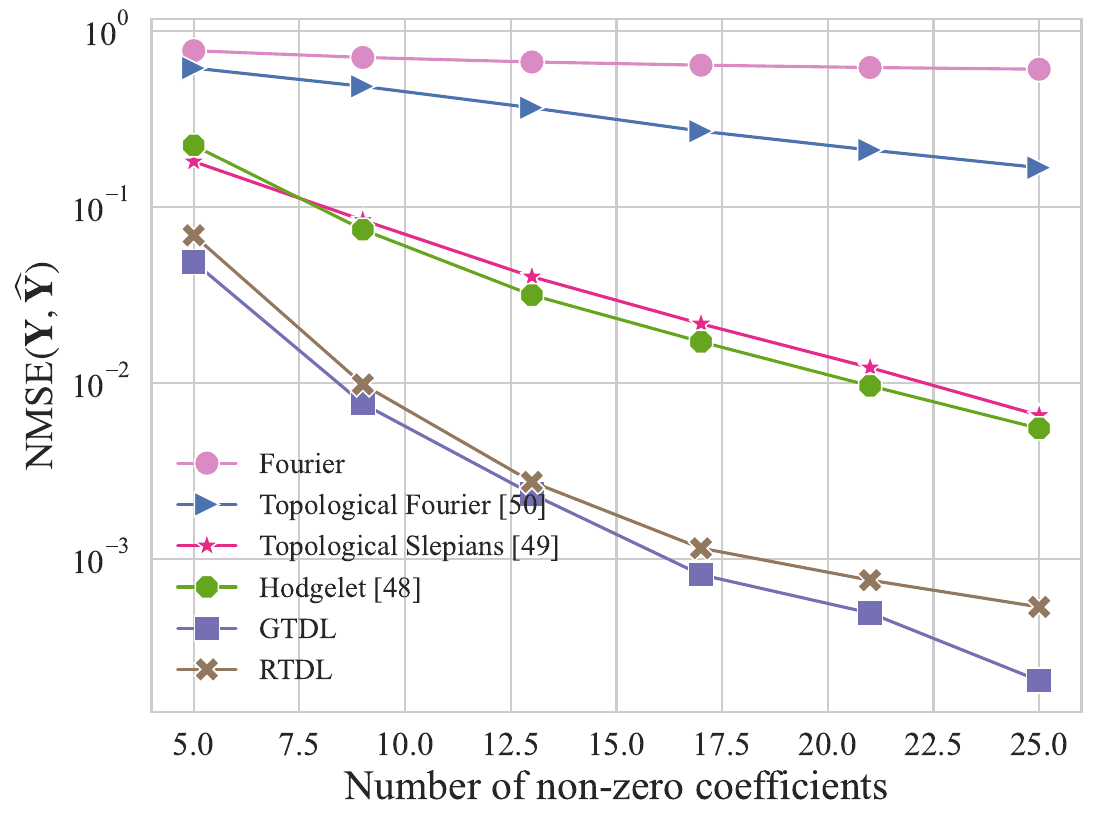}
    \caption{NMSE versus representation coefficients in a real-data scenario, considering different compression algorithms.
    }
    \label{fig:SignalReal}
\end{figure}

In this section, we present an empirical evaluation of our proposed topological dictionary learning framework using a real-world dataset. 
Specifically, we employ a dataset of real edge signals obtained from the German National Research and Education Network, which is operated by the DFN-Verein (DFN) \cite{orlowski2010sndlib}. In this real-world scenario, we analyze a second-order cell complex comprising $50$ vertices, $89$ edges, and $39$ potential polygons. The signals represent traffic data that have been aggregated on a daily basis during February 2005, with each measurement expressed in Mbit/sec and collected from each network link. As a result, we obtain a collection of edge signals denoted by $\mathbf{Y} \in \mathbb{R}^{89 \times 28}$. The dataset is partitioned into a training set $\mathbf{Y}_{tr}$ containing $21$ signals and a testing set $\mathbf{Y}_{te}$ comprising $7$ signals. Figure \ref{fig:SignalReal} illustrates the sparse representation capabilities of our algorithms in comparison with several analytic-dictionary-based methodologies available in the literaure, including: Topological Slepians \cite{battiloro2023topological}, Hodgelets \cite{roddenberry2022hodgelets}, Topological Fourier \cite{barbarossa2020topological}, and the classical discrete Fourier transform. As shown in Fig. \ref{fig:TopoReal}, the proposed GTDL and RTDL methods achieve a significant performance improvement compared to competing strategies. Furthermore, Fig. \ref{fig:TopoReal} illustrates the inferred second-order cell complexes derived from the application of the GTDL procedure, for different values of assumed sparsity $K_0$ equal to 13, 17, and 21. Interestingly, a consistent set of polygons (depicted in red) appears in all cases shown in Fig. \ref{fig:TopoReal}, highlighting their significance in representing the signal across different sparsity levels. These findings underscore the effectiveness of our approach in both signal representation and topology inference, thereby validating its potential for real-world applications where precise modeling of complex network structures is essential.

\section{Conclusions}
In this paper, we introduced a topological dictionary learning framework that exploits the structure of higher-order cell complexes to improve topological signal representation and infer the underlying domain topology. By leveraging Hodge theory, we embed topology into the design of an overcomplete dictionary, ensuring both topological and spectral localization properties. The learning process is formulated as a joint optimization problem over the topological domain, dictionary coefficients, and sparse signal representation, which is efficiently solved through iterative alternating algorithms. Extensive numerical experiments on both synthetic and real datasets demonstrate that the proposed methods yield substantial improvements in signal reconstruction compared to baseline techniques, while simultaneously providing valuable insights into the intrinsic topological structure of the data. Several promising research directions can be explored, including the learning of task-oriented directional topologies, scaling topology inference for large-scale systems, jointly processing multiple topological signals in the spectral domain, and integrating model-based deep learning techniques to enable deep unrolling of iterative topology inference algorithms.

\vspace{-.2cm}
\appendix
\label{appendix}

\subsection{Proof of Lemma \ref{lemma:qp_ref}} \label{appendixB}
The objective function of $\mathbf{P}_1$ can be recast as
\begin{align*}
 f(\mathbf{h}) &= \|\mathbf{Y - D(h, \overline{p}) \overline{S}}\|^2_F + \gamma \|\mathbf{h}\|^2_2 \\
&\hspace{-.4cm}=  \sum_{n=1}^{N} \sum_{m=1}^{T} [\mathbf{Y - D\overline{S}}]^2_{mn} + \gamma \|\mathbf{h}\|^2_2 \\
&\hspace{-.4cm}\overset{(a)}{=} \sum_{n=1}^{N} \sum_{m=1}^{T} \Bigg( [\mathbf{Y}]_{mn} - \sum_{i=1}^{M}  h_i^{(id)} \left[\mathbf{I}\overline{\mathbf{S}}_i\right]_{mn} \\
& \quad \: + \sum_{i=1}^{M}\sum_{j=1}^{J} h^{(u)}_{ij} \left[(\mathbf{L}^{(u)}_k)^j\overline{\mathbf{S}}_i\right]_{mn} \\
& \quad \: + \sum_{i=1}^{M}\sum_{j=1}^{J}  h^{(d)}_{ij} \left[(\mathbf{L}^{(d)}_k)^j \overline{\mathbf{S}}_i\right]_{mn} \Bigg)^2 + \gamma \|\mathbf{h}\|^2_2 \\
& \hspace{-.4cm}\overset{(b)}{=}  \sum_{n=1}^{N} \sum_{m=1}^{T} ([\mathbf{Y}]_{mn} - \mathbf{v}^T_{mn} \mathbf{h})^2 + \gamma \mathbf{h}^T \mathbf{h}\\
& \hspace{-.4cm}= \|\mathbf{Y}\|^2_F - \underbrace{2 \sum_{n=1}^{N} \sum_{m=1}^{T} [\mathbf{Y}]_{mn}\mathbf{v}^T_{mn}}_{\mathbf{r}} \mathbf{h} \: + \\
& \quad +\mathbf{h}^T \underbrace{\left( \sum_{n=1}^{N} \sum_{m=1}^{T} \mathbf{v}_{mn}\mathbf{v}^T_{mn} + \gamma \mathbf{I} \right)}_{\mathbf{Q}} \mathbf{h},
\end{align*}
where in $(a)$ we used (\ref{eq:dict_param}), and in $(b)$ we exploited the definitions in (\ref{eq:final_aux})-(\ref{eq:aux_P}). Clearly, $f(\mathbf{h})$ is a strongly convex quadratic function with $\mathbf{Q}$ being positive definite. 

\begin{figure}[t]
    \centering
    \includegraphics[width=1.\linewidth]{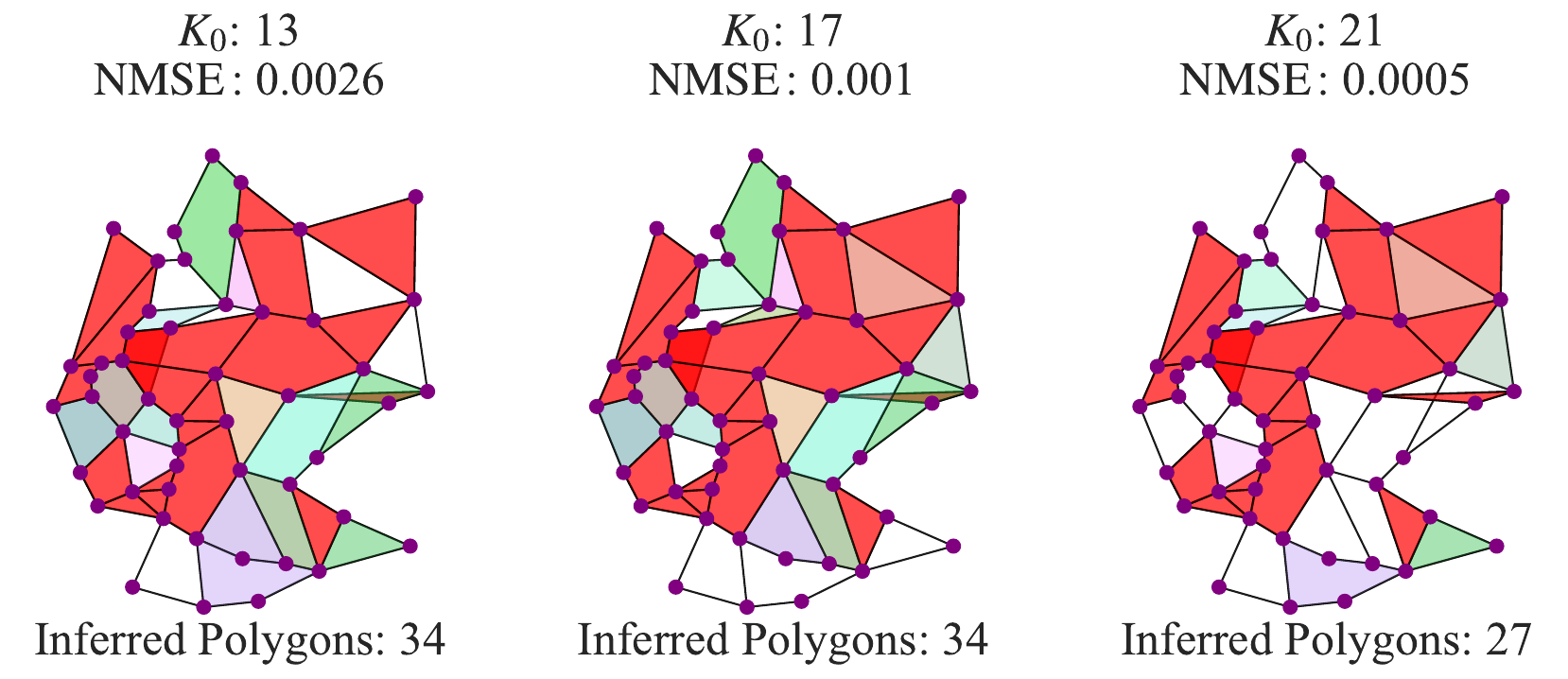}
    \caption{Examples of topologies inferred by the GTDL algorithm applied to real data from the DFN network.
    }
    \label{fig:TopoReal}
\end{figure}

Finally, we reformulate constraints $(b)$ and $(c)$ in $\mathbf{P}_1$ as affine functions. Constraint $(b)$ is equivalent to 
\begin{equation}
    0 \leq \hat{g}_i(l) \leq d, \quad \forall \, i = 1, \ldots, M, \quad l = 1, \ldots, N, 
\end{equation}
where $\hat{g}_i$ represents the kernel parameterized as in \eqref{eq:kernel_scalar} corresponding to $\mathbf{D}_i$. Defining  $\hat{\mathbf{g}}_i = \begin{bmatrix} \hat{g}_i(1), \ldots, \hat{g}_i(N) \end{bmatrix}^T$
and leveraging \eqref{eq:vand_constr1}–\eqref{eq:vand_constr3}, we obtain 
\[
    \hat{\mathbf{g}}_i = \mathbf{F} \mathbf{h}_i,
\]
with $\mathbf{h}_i\in \mathbb{R}^{2J+1}$ denoting the portion of vector $\mathbf{h}$ associated with the $i$-th dictionary. Consequently, for all $i = 1, \ldots, M$, the constraint \eqref{eq:spec_const1}, i.e., $(b)$ in $\mathbf{P}_1$, can be compactly cast as:
\begin{equation}
    \mathbf{0} \leq \mathbf{I}_M \otimes \mathbf{F} \mathbf{h} \leq d \mathbf{1}.
    \label{eq:spec_const1_compact}
\end{equation}

By following a similar reasoning, constraint $(c)$ can be reformulated as:
\begin{equation}
    (d - \varepsilon) \, \underline{\mathbf{1}} \leq (\underline{\mathbf{1}}^T \otimes \mathbf{F}) \, \mathbf{h} \leq (d + \varepsilon) \, \underline{\mathbf{1}}.
    \label{eq:spec_const2_compact}
\end{equation}
This concludes the proof of Lemma 2.

\balance
\bibliographystyle{IEEEbib}
\bibliography{adj_biblio}

\end{document}